\documentclass{emulateapj}

\usepackage{amsmath}
\usepackage{booktabs}
\usepackage{graphicx}
\usepackage{natbib}
\usepackage{hyperref}

\newcommand{\mf}{\langle F \rangle}
\newcommand{\dperp}{\langle d_{\perp} \rangle}

\newcommand{\HI}{\textrm{H}\textsc{i}}
\newcommand{\Msun}{M_{\odot}}

\newcommand{\Nmap}{N_{\rm map}}
\newcommand{\Npix}{N_{\rm pix}}

\newcommand{\SNR}{\mathrm{SNR}}

\newcommand{\tr}{\mathrm{tr}}
\renewcommand{\vec}[1]{\mathbf{#1}}
\newcommand{\vb}{\vec b}
\newcommand{\vd}{\vec d}
\newcommand{\vk}{\vec k}
\newcommand{\vm}{\vec m}
\newcommand{\vn}{\vec n}
\newcommand{\vs}{\vec s}
\newcommand{\vsh}{\hat{\vec s}}
\newcommand{\vw}{\vec w}
\newcommand{\vx}{\vec x}

\newcommand{\vA}{\vec A}

\newcommand{\vI}{\vec I}
\newcommand{\vG}{\vec G}
\newcommand{\vL}{\vec L}
\newcommand{\vM}{\vec M}
\newcommand{\vN}{\vec N}
\newcommand{\vS}{\vec S}

\newcommand{\vSmm}{\vec S_{mm}}
\newcommand{\vSmp}{\vec S_{mp}}
\newcommand{\vSpm}{\vec S_{pm}}
\newcommand{\vSpp}{\vec S_{pp}}

\begin{document}

\title{Protocluster Discovery in Tomographic Ly$\alpha$ Forest Flux Maps}

\author{Casey W. Stark$^{1}$, Martin White$^{1, 2, 3}$, Khee-Gan Lee$^{4}$,
Joseph F. Hennawi$^{4}$ \\
$^1$ Department of Astronomy, University of California, Berkeley, CA 94720,
USA \\
$^2$ Department of Physics, University of California, Berkeley, CA 94720, USA \\
$^3$ Lawrence Berkeley National Laboratory, 1 Cyclotron Road, Berkeley, CA
93720, USA \\
$^4$ Max Planck Institute for Astronomy, K\"{o}nigstuhl 17, D-69117 Heidelberg,
West Germany \\
}

\begin{abstract}
We present a new method of finding protoclusters using tomographic maps of
Ly$\alpha$ Forest flux. We review our method of creating tomographic flux maps
and discuss our new high performance implementation, which makes large
reconstructions computationally feasible. Using a large $N$-body simulation,
we illustrate how protoclusters create large-scale flux decrements, roughly
$10 \, h^{-1}$Mpc across, and how we can use this signal to find them in flux
maps.
We test the performance of our protocluster finding method by running it on
the ideal, noiseless map and tomographic reconstructions from mock surveys,
and comparing to the halo catalog.
Using the noiseless map, we find protocluster candidates with about
90\% purity, and recover about 75\% of the protoclusters that form
massive clusters ($> 3 \times 10^{14} \, h^{-1} M_{\odot}$).
We construct mock surveys similar to the ongoing COSMOS Lyman-Alpha Mapping And
Tomography Observations (CLAMATO) survey.
While the existing data has an average sightline separation of
$2.3 \, h^{-1}$Mpc, we test separations of 2 -- $6 \, h^{-1}$Mpc
to see what can be tolerated for our application.
Using reconstructed maps from small separation mock surveys, the protocluster
candidate purity and completeness are very close to what was found in the
noiseless case.
As the sightline separation increases, the purity and completeness
decrease, although they remain much higher than we initially expected.
We extended our test cases to mock surveys with an average separation of
$15 \, h^{-1}$Mpc, meant to reproduce high source density areas of the
BOSS survey. We find that even with such a large sightline separation, the
method can still be used to find some of the largest protoclusters.
\end{abstract}

\maketitle

\section{Introduction}

Galaxy clusters are the largest and most massive gravitationally-bound
structures in the Universe, the endpoint of a long process of hierarchical
structure formation. Due to their large mass, deep potential wells and dynamic
formation histories they are important laboratories for studying galaxy
evolution, plasma physics, and our models of gravity and cosmology
\citep{fabian_1994, krastov_and_borgani_2012, feretti_et_al_2012}. Despite keen
interest in how clusters form, the study of early cluster formation, at high
$z$, is observationally limited: clusters are rare and surveying large volumes
is expensive. Indeed, the total comoving volume of even the largest surveys for
distant galaxies at $z \sim 2$ -- 3 \citep[e.g.\ KBSS,][]{rudie_et_al_2012} is
only $\sim 10^{7} \, \mathrm{Mpc}^3$, which would barely contain a single rich
cluster locally. In the past decade small samples of protoclusters have been
compiled \citep[see e.g.][for a recent compilation]{chiang_et_al_2013,
chiang_et_al_2014} but important questions regarding the formation of clusters
and the evolutionary tracks of member galaxies remain unresolved
\citep[e.g.][]{peterson_and_fabian_2006, dolag_et_al_2009, martizzi_et_al_2014}.
There has been progress in the theoretical understanding of cluster formation
through the use of $N$-body simulations and semi-analytic galaxy formation
models \citep{baugh_2006, benson_and_bower_2010, benson_2012} and hydrodynamical
simulations \citep{sijacki_and_springel_2006, mccarthy_et_al_2010,
yang_et_al_2012, skory_et_al_2013, vogelsberger_et_al_2013, genel_et_al_2014},
although it is a notoriously difficult problem to predict member galaxy
properties from first principles. It is an area of ongoing research to validate
and extend the numerous assumptions and subgrid recipes which are made in these
works.

With the advent of large surveys in the optical and near-IR
\citep{postman_et_al_1996, kochanek_et_al_2003, gladders_yee_2005,
koester_et_al_2007, wilson_et_al_2009, muzzin_et_al_2009, hao_et_al_2010,
szabo_et_al_2011, murphy_et_al_2012, rykoff_et_al_2014, bleem_et_al_2014},
sub-mm \citep{marriage_et_al_2011, reichardt_et_al_2013, planck_2013_XXIX} and
X-ray \citep{ebeling_et_al_2000, bohringer_et_al_2004, burenin_et_al_2007,
pierre_et_al_2006, finoguenov_et_al_2007} bands we now have large samples of
clusters, with a tail extending beyond $z \simeq 1$ -- 2.
These surveys leverage the fact that `mature'
clusters contain large overdensities of (typically red) galaxies and a hot
intracluster medium. Protoclusters, at $z = 2$ or earlier, lack these signatures
making them more difficult to find and study. At the time of writing only a few
tens of protocluster candidates are known at $z > 2$
\citep[see][]{chiang_et_al_2013, chiang_et_al_2014, finley_et_al_2014,
cucciati_et_al_2014},
and most candidates were found via the the signpost technique,
i.e.\ using a radio-galaxy, Ly$\alpha$ blob, or another source as a marker.

Assuming a mean interior density of $200$ times the background density, the
linear size of the mean-density region from which material is accreted into a
halo should be about 5 -- 6 times the virial radius of the final halo.
For protoclusters this can be up to $10 \, h^{-1}$Mpc, i.e.\ we expect that the
$z \sim 2$ progenitors of massive clusters should lie in overdense regions many
(comoving) Mpc in radius. This expectation is born out of numerical simulations
\citep[e.g.][]{chiang_et_al_2013} which also show that the most massive clusters
today form not from the most overdense regions at high $z$ but from large,
possibly only moderately overdense regions \citep{angulo_et_al_2012}.
The progenitor regions of massive low-$z$ clusters should thus be identifiable
in relatively low-resolution large-scale structure maps of the high-$z$
Universe. Systematic searches in large, deep, galaxy redshift surveys or
multi-band photometric surveys are one promising way to find protoclusters
\citep[e.g.][]{chiang_et_al_2014, diener_et_al_2014, yuan_et_al_2014}, although
projection effects pose a challenging problem. Spectroscopic surveys with
sufficient sampling of Mpc-scales take care of this problem, although redshift
errors can still be significant and covering large volumes with such high
resolution is prohibitively expensive.

An alternative is tomographic mapping using Ly$\alpha$ absorption from neutral
Hydrogen in the intergalactic medium (IGM) \citep{caucci_et_al_2008,
cai_et_al_2014, lee_et_al_2014a}. \citet{lee_et_al_2014a} demonstrated that IGM
tomography allows large volumes of the Universe to be efficiently searched for
protoclusters in the $z \simeq 2$ -- 3 range using existing facilities. By
targeting star-forming Lyman-break galaxies (LBGs) as well as quasars at
$g \gtrsim 24.5$, at signal-to-noise ratios achievable with existing facilities,
we can obtain hundreds of sightlines per deg$^2$. This sightline density
corresponds to average spacings of several Mpc, which is also the correlation
scale of the Ly$\alpha$ forest.
By sampling the IGM absorption along and across sightlines with Mpc
spacing, we are able to tomographically reconstruct the 3D Ly$\alpha$ forest
flux field. These tomographic maps have a resolution similar to the average
transverse sightline spacing and naturally avoid projection effects or redshift
errors. In \citet{lee_et_al_2014c}, we constructed a tomographic IGM map using
24 LBG spectra with an average sightline separation $\dperp = 2.3 \, h^{-1}$Mpc,
obtained with two 2-hour exposures on Keck LRIS. These observations made up the
pilot data of the COSMOS Lyman-Alpha Mapping And Tomography Observations
(CLAMATO) survey, which we plan to extend to cover 1 deg$^2$.
These observations will result in a tomographic map with a volume of roughly $70
\times 70 \times 230 \, h^{-1}$Mpc and $2 \, h^{-1}$Mpc resolution. Such a map
will provide an unprecedented view of the intergalactic medium and provide a
large volume to search for protoclusters.

Given the diversity of protoclusters, the ability to construct large samples is
important if we are to draw robust conclusions about cluster formation. Optical,
sub-mm and X-ray facilities could then be used to follow up the most promising
candidates looking for galaxy over densities, Compton decrements or faint,
diffuse X-ray emission. The HETDEX \citep{hill_et_al_2004} and Subaru Prime
Focus \citep{takada_et_al_2014} spectrographs would be particularly
powerful for following up such candidates in the optical. As we shall show
later, the most massive progenitors of the most massive clusters today ($M > 3
\times 10^{14} \, h^{-1} \Msun$) can reach the rich group scale ($M \sim 3
\times 10^{13} \, h^{-1} \Msun$) before $z \sim 2$. Such structures could well
have observable galaxy overdensities and a hot gas component at early times.

The outline of the paper is as follows. In Section~\ref{sec:sim} we introduce
the numerical simulation we use to study protocluster recovery using IGM
tomography. The properties of the protoclusters are discussed in
Section~\ref{sec:pc}, while Section~\ref{sec:recon} describes the algorithm we
use to make maps from our mock observations. In Section~\ref{sec:pcid}, we
describe a simple method for identifying protocluster candidates in the
reconstructed maps. In Section~\ref{sec:surveys}, we synthesize mock surveys
from our simulation, and test how protocluster identification depends on survey
parameters. Finally, we provide a summary of our findings in
Section~\ref{sec:conclusions}.

\section{Simulations}
\label{sec:sim}

In order to validate our protocluster search strategy, and to study the purity
and completeness of the sample we obtain, we make use of cosmological $N$-body
simulations. We require a simulation which simultaneously covers a large
cosmological volume (to have a statistically fair sample of the rare clusters
and protocluster regions) while having a sufficiently small inter-particle
spacing to model transmission in the IGM. The requirements are sufficiently
demanding that we have used a pure $N$-body simulation, augmented with the
fluctuating Gunn-Peterson approximation \citep[FGPA;][]{petitjean_et_al_1995,
croft_et_al_1998, meiksin_and_white_2001, meiksin_2009}. This simulation was
also used in \citet{lee_et_al_2014c}.

\subsection{N-body simulation}

Our simulation employed $2560^3$ equal mass ($8.6 \times 10^7 \, h^{-1}
M_\odot$) particles in a $256 \, h^{-1}$Mpc periodic, cubical box leading to a
mean inter-particle spacing of $100 \, h^{-1}$kpc. This is sufficient to model
the large-scale features in the IGM at $z \simeq 2$ -- 3 using the FGPA
\citep{meiksin_and_white_2001, rorai_et_al_2013} and more than sufficient to
find clusters at $z = 0$. The assumed cosmology was of the flat $\Lambda$CDM
family, with $\Omega_{\rm m} \simeq 0.31$, $\Omega_{\rm b} h^2 \simeq 0.022$, $h
= 0.6777$, $n_s = 0.9611$, and $\sigma_8 = 0.83$, in agreement with
\citet{planck_2013_XVI}. The initial conditions were generated using second-
order Lagrangian perturbation theory at $z_{\rm ic} = 150$, when the rms
particle displacement was 40 per cent of the mean inter-particle spacing. The
particle positions and velocities were advanced to $z = 0$, using a TreePM code
\citep{white_2002} assuming a spline-softened force with a Plummer equivalent
smoothing length of $3 \, h^{-1}$kpc. This TreePM code has been compared to a
number of other codes and shown to perform well for such simulations
\citep{heitmann_et_al_2008}.

\subsection{Halo catalogs}

At $z = 0$ and $z = 2.5$, we generated halo catalogs using a friends-of-friends
\citep[FoF;][]{DEFW} algorithm with a linking length $b = 0.168$. This algorithm
partitions particles into groups bounded approximately by isodensity contours of
roughly 100 times the mean density \citep[e.g.][and references
therein]{lacey_cole_1994, white_2001}. Since we focus only on the most massive
objects in our simulations, FoF halos are sufficient --- more sophisticated halo
finding methods will recover more detailed halo and subhalo properties, but with
increased complexity and computational cost\footnote{For a recent review and
comparison of halo finding methods see \citet{knebe_et_al_2011}.}.

\subsection{Ly$\alpha$ flux field}

For the output at $z = 2.5$, we also generated mock Ly$\alpha$ forest spectra
on a $2560^3$ grid with the FGPA. This approximation makes use of the fact that
adiabatic cooling of the gas in the presence of a photoionizing ultraviolet
background leads to a tight density-temperature relation in the low density gas
responsible for the Ly$\alpha$ forest seen in absorption against bright objects
\citep{gnedin_hui_1998, meiksin_2009}. The approximation has been shown to match
more detailed hydrodynamical computations at the ten percent level
\citep{meiksin_and_white_2001, viel_et_al_2002, mcdonald_2003, viel_et_al_2006},
and is certainly sufficient for our purposes.

The dark matter particle positions and velocities were deposited onto the grid
using CIC interpolation \citep{hockney_and_eastwood_1988}. We then Gaussian
filtered the density and velocity on the grid in order to approximate the
pressure smoothing which affects the gas density. We assumed an IGM temperature
at mean density $T_0 = 2 \times 10^4 \, $K, which gives a filtering scale of
about $100 \, h^{-1}$kpc at the redshifts of interest here
\citep[e.g.][]{gnedin_hui_1998, viel_et_al_2002, white_et_al_2010,
rorai_et_al_2013}.
Our results are largely insensitive to the details of this pressure smoothing
procedure, since we are probing fluctuations on much larger scales (Mpc). We set
the temperature according to the density-temperature relation $T = T_0 (\rho /
\bar \rho)^{\gamma - 1}$, with a standard choice for the equation of state
parameter $\gamma = 1.6$ \citep{lee_et_al_2014b}. We compute the optical depth
to \HI\ Ly$\alpha$ scattering $\tau$ and the transmitted flux $F = e^{-\tau}$
assuming the \HI\ density is proportional to the ratio of the recombination and
photoionization rates $n_{\HI} \propto \rho^2 T^{-0.7} \Gamma^{-1}$ and that
the line profile is a Doppler profile, and we normalize the optical depth such
that the mean flux $\langle F \rangle = 0.8$, matching the recent observational
result in \citet{becker_et_al_2013} for this redshift. This scheme ignores
several phenomena that could affect the Ly$\alpha$ forest including
spatial fluctuations in the temperature of the IGM due to reionization
inhomogeneities, spatial fluctuations in the ultraviolet background due to the
shot noise of sources, and galactic outflows. Fortunately, at the Mpc scale,
the effects of galactic outflows and temperature fluctuations on flux should be
rather small, while we expect the ultraviolet background to fluctuate on scales
of several hundred Mpc
\citep{mcdonald_et_al_2005, greig_et_al_2014, pontzen_2014, gontcho_et_al_2014}.
In the remainder of the paper, when we refer to flux,
we mean the Ly$\alpha$ forest transmitted flux fraction perturbation
$\delta_F = F / \langle F \rangle - 1$.

The final products we use from the simulation, then, are the halo catalogs at $z
= 0$ and 2.5, including the positions of the particles within those halos at $z
= 2.5$, and 3D grids of density and flux. We begin by studying the relationship
between this ideal flux field and the halos and protoclusters. In Section
\ref{sec:surveys}, we will look at the impact of finite sightline density,
resolution and noise on the recovery of the flux field.

\section{Protoclusters in density and Ly$\alpha$ forest flux}
\label{sec:pc}

\begin{figure}
  \begin{center}
    \resizebox{\columnwidth}{!}{\includegraphics{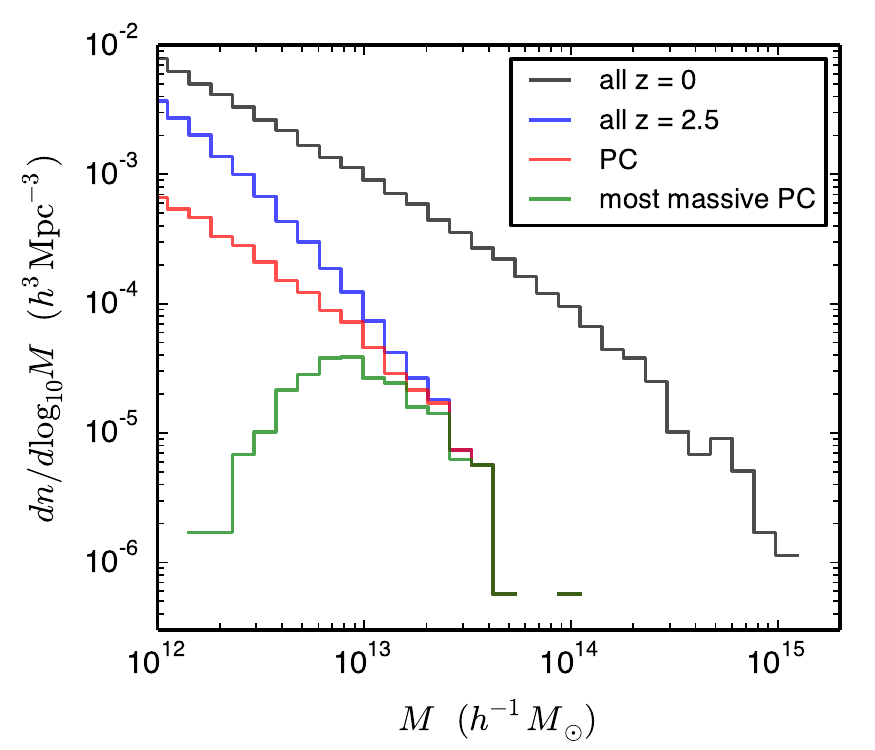}}
  \end{center}
  \caption{The halo mass functions for all halos at $z = 0$ (black), all halos
  at $z = 2.5$ (blue), protocluster halos (red), and the most massive halos
  in each protocluster (green). The massive end of the high redshift mass
  function is dominated by protocluster halos. The most massive halo in a
  protocluster is typically $10^{13} \, h^{-1} \Msun$ at this redshift.}
  \label{fig:mass_funcs}
\end{figure}

\begin{figure*}
  \begin{center}
    \resizebox{\textwidth}{!}{\includegraphics{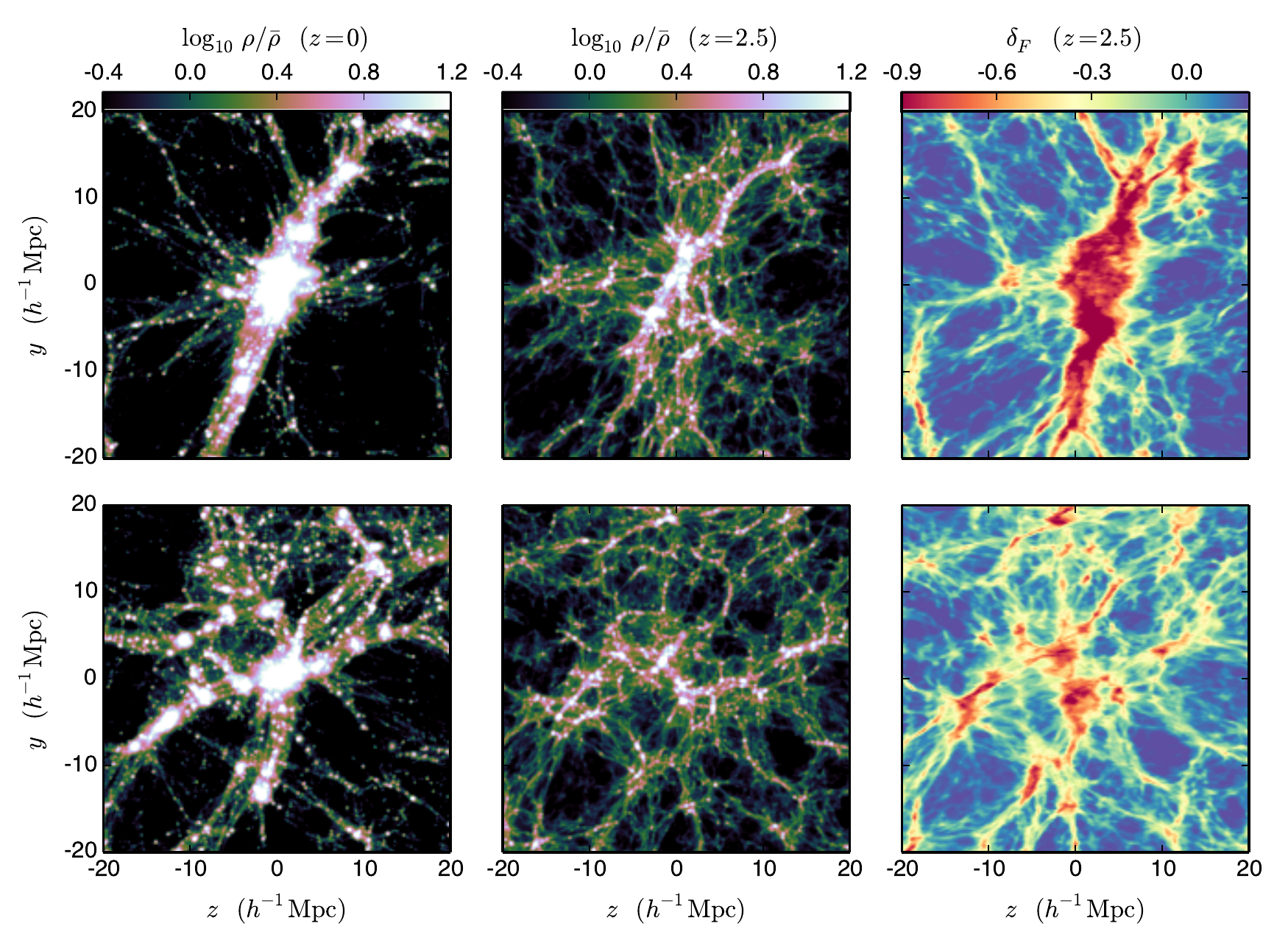}}
  \end{center}
  \caption{Slices through the density and flux fields centered on two
protoclusters. The line-of-sight direction is horizontal. The upper row shows
a cluster/protocluster which is easily found by our method,
while the lower row shows a more problematic case. The upper row cluster has
a mass $M = 9 \times 10^{14} \, h^{-1} \Msun$, while the lower row cluster has
a mass $M = 3 \times 10^{14} \, h^{-1} \Msun$. We chose these sample
protoclusters based on the $\delta_F$ value at the protocluster center of mass
(COM), where the top protocluster has the smallest $\delta_F$ value, and the
bottom protocluster has the largest. The slices are $40 \times 40 \, h^{-1}$Mpc
on a side and $5 \, h^{-1}$Mpc thick. In each row, the color scale shows the log
overdensity or flux perturbation: (Left) the $z = 0$ density (centered on the
cluster COM), (Middle) the $z = 2.5$ density (centered on the protocluster COM),
(Right) the $z = 2.5$ flux perturbation, $\delta_F$. Note that overdense
regions correspond to regions of increased absorption, or more negative
$\delta_F$, and that correlation is quite strong on these scales. The small
differences in the $z = 2.5$ density and flux fields are due to the density
is shown in real-space while the flux is in redshift-space.}
  \label{fig:progen_slices}
\end{figure*}

The boundary between a rich group and a cluster is somewhat arbitrary, but we
shall define a cluster at $z = 0$ as a halo with a FoF mass larger than $10^{14}
\, h^{-1} M_\odot$. We have 425 halos above this mass in the simulation at $z =
0$ and these will form our sample. A protocluster is the high-redshift
progenitor of such massive halos, but due to the hierarchical process by which
halos form, there is some ambiguity as to what constitutes the progenitor. At $z
\simeq 2$ -- 3, the mass which will eventually lie within the $z = 0$ halo is
spread among several relatively large progenitor halos and in the nearby IGM,
spread over tens of (comoving) Mpc. We tracked the cluster progenitor halos by
finding all halos at $z = 2.5$ that contributed half or more of their mass to
the resulting cluster. We show the mass functions of all $z = 2.5$ halos, of
protocluster halos, and of the most massive halo in each protocluster in
Figure~\ref{fig:mass_funcs}. The high-mass end of the mass function is dominated
by the halos that form clusters, but the protocluster halos do not make up all
of the high mass halos. Protocluster halos only make up about half of the halos
near $10^{13} \, h^{-1} \Msun$ for instance. We found that the most massive
progenitor halo is typically about $10^{13} \, h^{-1} \Msun$, with more massive
clusters having more massive progenitor halos on average. Only the most massive
such halos are likely to host a hot, X-ray emitting ICM or be found as
significant overdensities of galaxies. We also computed the second moment of the
progenitor halo positions
$\sqrt{ [\sum_i m_i ( \vx_i - \bar \vx )^2] / [\sum_i m_i] }$, where $\bar \vx$
is the average position and $m_i$ and $\vx_i$ are the
halo masses and centers, as done in \citet{chiang_et_al_2013} to confirm the
extent of the halos they found at this redshift. We found that the progenitor
halos are spread over 4 -- $8 \, h^{-1}$Mpc, in good agreement with their values
at $z = 2$ -- 3. However, in contrast to \citet{chiang_et_al_2013}, we are
interested less in the progenitor halos and more in the large-scale overdense
region from which the mass of the cluster will be assembled.

In order to define the protocluster center, we tracked particles that form the
core of the $z = 0$ cluster back to $z = 2.5$, and computed their center of mass
(COM). The choice of particles that constitute the `core' of the cluster is
arbitrary, but the exact choice of particles does not matter as long as the
resulting COM does not change significantly. We chose to select the particles
within $200 \, h^{-1}$kpc from the most bound (densest) cluster particle at $z
= 0$. We refer to this collection of particles that makes up the cluster core as
the $N$-densest particles. We found that changing the cutoff radius from 100 to
$500 \, h^{-1}$kpc results in small changes to the protocluster center, on the
level of $100 \, h^{-1}$kpc, which is negligible for objects spanning several
Mpc. Inspired by \citet{chiang_et_al_2013}, we define the protocluster radius
$r_{\rm pc}$ as the radius of a sphere, centered on the protocluster center,
enclosing 50 percent of the particles which belong to the halo at $z = 0$.
We found the expected trend that more massive clusters have larger protocluster
sizes. The $10^{\rm th}$ percentile radius is $3.3 \, h^{-1}$Mpc, the
$50^{\rm th}$ percentile is $4.1 \, h^{-1}$Mpc, and the $90^{\rm th}$ percentile
is $5.4 \, h^{-1}$Mpc. The largest half-mass radius we found in the simulation
is $8.9 \, h^{-1}$Mpc, and this protocluster forms a $10^{15} \, h^{-1} \Msun$
cluster. This, in combination with the moment of the progenitor
halo positions, gives us good reason to believe that protoclusters will stand
out on scales of $\sim 4 \, h^{-1}$Mpc at this redshift.

We show two examples of protoclusters, as seen in density and Ly$\alpha$ forest
flux, in Figure~\ref{fig:progen_slices}. The upper row shows a protocluster with
a large coherent structure which will be easily found by our method, while the
lower row shows a case where the protocluster is spread out and will prove much
more difficult to find. The upper row cluster has a mass
$M = 9 \times 10^{14} \, h^{-1} \Msun$, while the lower row cluster has a mass
$M = 3 \times 10^{14} \, h^{-1} \Msun$.
From left to right, we show the $z = 0$ density, $z = 2.5$ density, and the
flux in a slice $40 \, h^{-1}$Mpc across and $5 \, h^{-1}$Mpc thick.
Due to the physics of the IGM, the flux is tightly correlated with the matter
density on large scales, with overdense regions leading to more absorption (low
flux). In the protocluster in the upper row, the progenitor halos that merge to
form the cluster can be easily seen in the middle column and lead to a large,
coherent flux decrement in the right column. The flux decrement in the lower row
is still visible, but it is not as pronounced, because the halos making up the
protocluster are more diffuse. We compared the progenitor halos of these
clusters and found that at fixed mass, the protocluster in the upper row has
three times as many halos and that the most massive halo is twice as massive,
indicating that the upper row cluster forms earlier. The most massive progenitor
halo in the upper row cluster has a mass of $3 \times 10^{13} \, h^{-1} \Msun$
--- a typical rich group mass --- meaning that it should be easier to follow up
at high redshift. Overall, we found that 40\% of the protoclusters contain a
halo with a mass $M > 10^{13} \, h^{-1} \Msun$.

\begin{figure}
  \begin{center}
    \resizebox{\columnwidth}{!}{\includegraphics{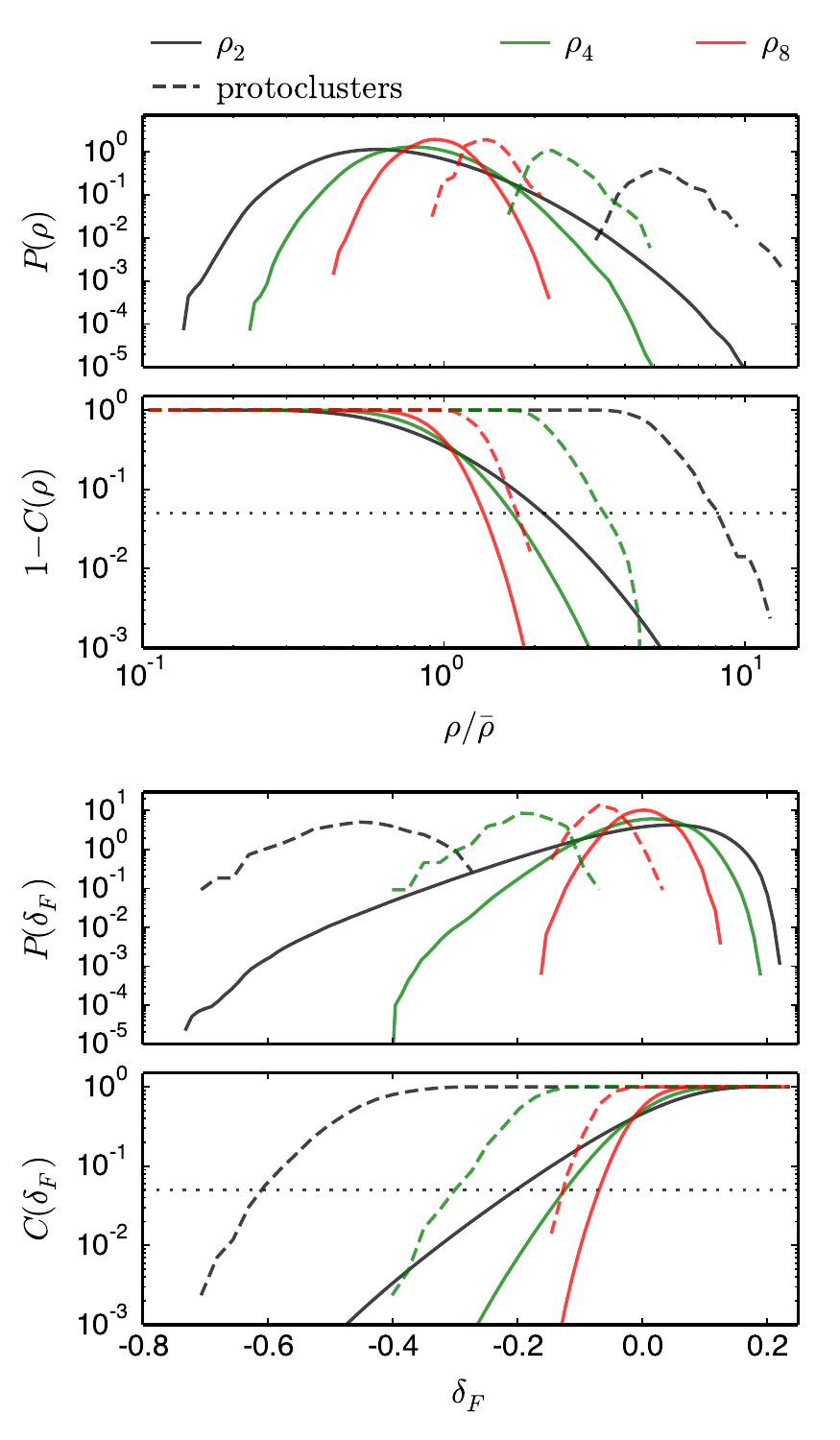}}
  \end{center}
\caption{Distributions of the matter density and flux, smoothed with Gaussians
of $\sigma = 2$, 4, and $8 \, h^{-1}$Mpc (labeled $\rho_2$, $\rho_4$, and
$\rho_8$ with the broadest distributions having the smallest $\sigma$). Solid
lines show the PDF for the entire volume while the dashed lines indicate the
densities or fluxes at the protocluster positions. The top two panels show the
matter density PDF, $P(\rho)$, and the cumulative distribution, $C(\rho)$,
plotted as $1 - C$ to highlight regions of high density. The horizontal dotted
black line shows the $95^{\rm th}$ percentile. We see that protoclusters
preferentially lie in the highest density regions of the density field, smoothed
on Mpc scales. The lower two panels show the PDF and cumulative distribution for
the flux perturbation, $\delta_F$. We see that protoclusters preferentially lie
in the negative tails of the distribution.}
  \label{fig:field_pc_dists}
\end{figure}

Not surprisingly, all of the protocluster regions lie on the high-density tail
of the density distribution. We smoothed the density field with Gaussian filters
of scales 2, 4, and $8 \, h^{-1}$Mpc (labeled $\rho_2$, $\rho_4$, and $\rho_8$
respectively) and compared the distributions of the full field and the
protoclusters. We smooth the fields for two reasons: to mimic the characteristic
resolution of our tomographic maps and because protoclusters should stand out
most on scales of several Mpc. The top panel of Figure~\ref{fig:field_pc_dists}
shows the probability density function of the density $p(\rho)$ for random
positions (solid) and for the protocluster regions (dashed). The majority of the
protoclusters have densities exceeding the $95^{\rm th}$ percentile of the
density distribution. This is clearer in the second panel showing the cumulative
distribution $C(\rho)$, plotted as log-scaled $1 - C$ to highlight the
high-density tail. Here, it is easy to see the $95^{\rm th}$ percentile density for
the field, and compare to the protocluster distribution. Regardless of the
smoothing scale, nearly all protoclusters have densities in the $95^{\rm th}$
percentile tail. In the bottom two panels, we show the probability density and
cumulative distribution of the flux. Since the large-scale flux is
so tightly correlated with the density, we find that the majority of
protoclusters similarly lie in the low-flux tail of the distribution.
Protoclusters can thus be found quite efficiently by searching for large-scale
flux decrements \citep[see also][]{cai_et_al_2014}. In 1D, large-scale flux
decrements can also be created by damped Ly$\alpha$ systems (DLA)
\citep{meiksin_2009}. However, DLAs have physical
extents of $< 100 \, h^{-1}$kpc, much smaller than our transverse scales,
which make it very unlikely for DLAs to contaminate several nearby sightlines
at the same redshift.

\begin{figure}
  \begin{center}
    \resizebox{\columnwidth}{!}{\includegraphics{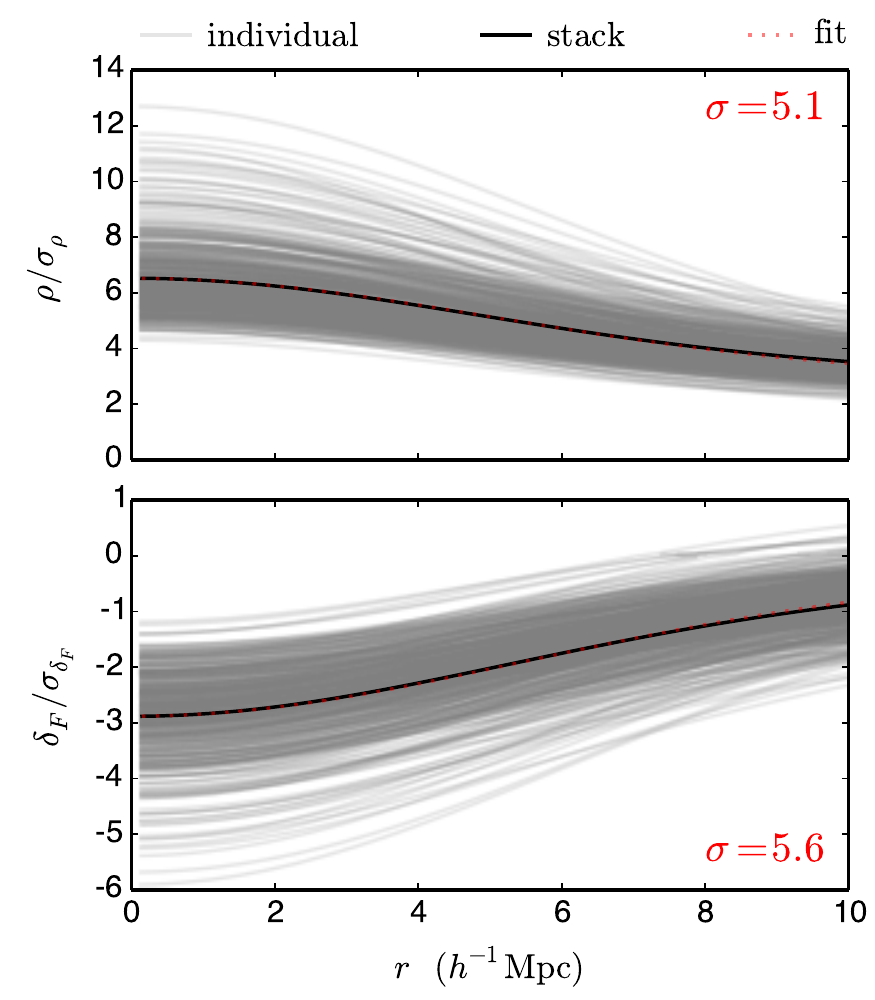}}
  \end{center}
\caption{The radial density and flux profiles of the protocluster regions
(from the $4 \, h^{-1}$Mpc smoothed fields). We show the individual profiles in
grey, the average profiles in black, and Gaussian fits in dotted red. (Top)
the density profiles. We plot $\rho / \sigma_\rho$, where $\sigma_\rho$ is the
standard deviation of the field, since we are interested in how extreme the
protocluster regions are. (Bottom) repeated with flux. The fit Gaussian scale
$\sigma$ is annotated in red.}
  \label{fig:pc_profiles}
\end{figure}

The radial profiles of the protocluster in density and flux are shown in
Figure~\ref{fig:pc_profiles}. These profiles were constructed by radially
binning the $4 \, h^{-1}$Mpc smoothed fields, from the center of each
protocluster (grey lines) and by stacking all protocluster profiles (black
lines). Again, we use the smoothed fields to mimic the tomographic map
resolution and to highlight protocluster scales. On the y-axis in both panels,
we plot the standard-deviation normalized values (where we use the standard
deviation of the smoothed field) to see how much protocluster profiles stand out
relative to other fluctuations at this scale. The overdensity and flux decrement
near the center is significant. We found that the profiles are even more
pronounced in the $2 \, h^{-1}$Mpc smoothed fields, while in the $8 \,
h^{-1}$Mpc smoothed fields, the profiles are shallow, and do not stand out
significantly in the center. This indicates that smoothing at a scale of $8 \,
h^{-1}$Mpc is likely too aggressive for our application. We fit Gaussian
profiles to the average density and flux profiles, and show the fits with dotted
red lines. We also annotated the fit Gaussian $\sigma$ values, which indicate
that the protoclusters are overdense/under-fluxed over several Mpc. These
results validate our strategy for finding protoclusters by looking for
large-scale flux decrements in the Ly$\alpha$ forest.

\begin{figure}
  \begin{center}
    \resizebox{\columnwidth}{!}{\includegraphics{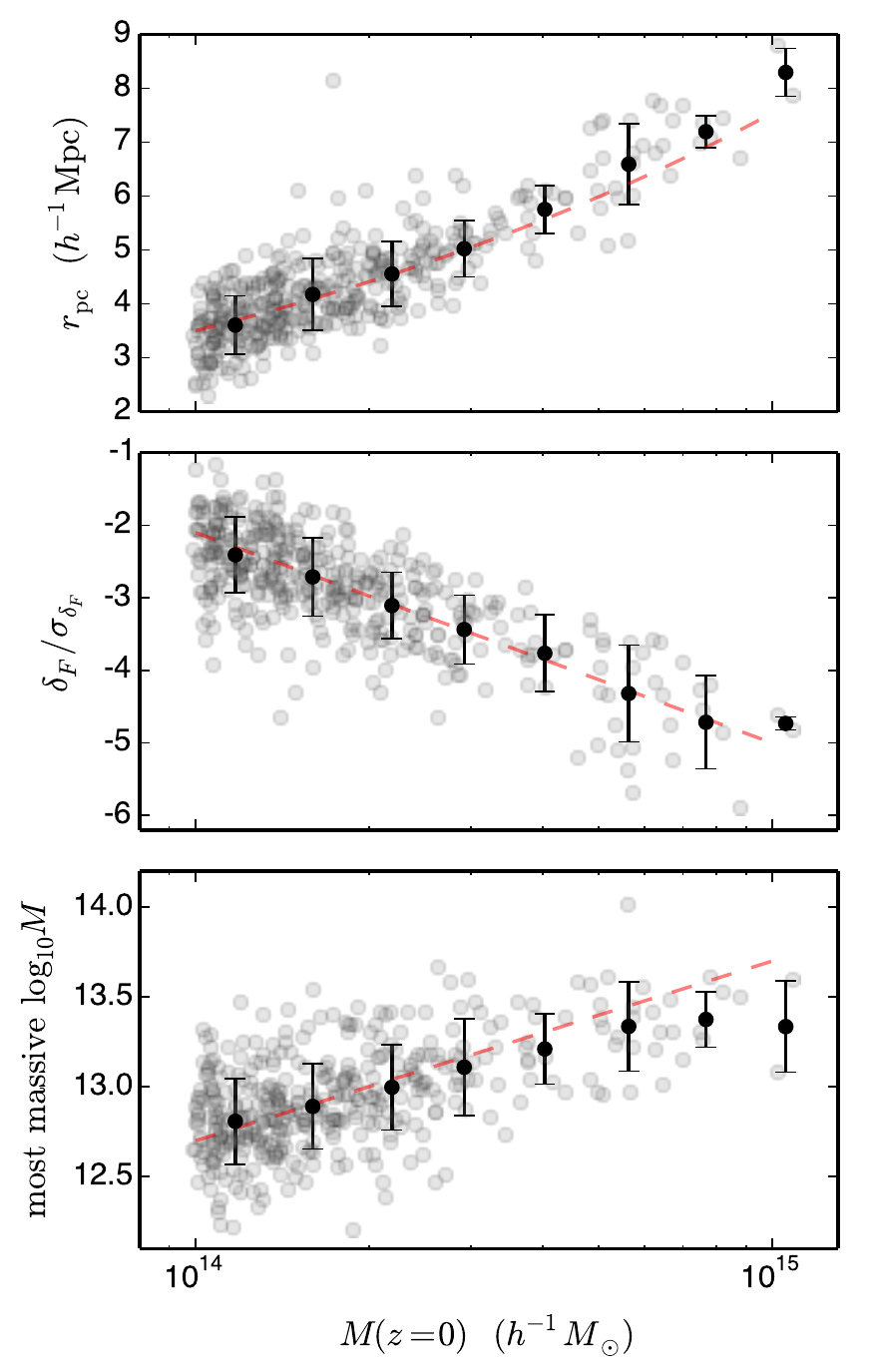}}
  \end{center}
\caption{Several protocluster properties vs.\ the resulting cluster mass
$M(z = 0)$. In each panel, we plot each protocluster as a gray dot, and the
$M(z = 0)$ binned result with std.\ dev.\ error bars in black. The red dashed
lines show an approximate scaling.
Top: the protocluster half-mass radius $r_{\rm pc}$. The red line is the
$r \propto M^{1/3}$ relation, which fits well.
Middle: the flux decrement $\delta_F / \sigma_{\delta_F}$, evaluated at the
protocluster center, smoothed with a 4 $h^{-1}$Mpc Gaussian.
We show $\delta_F / \sigma_{\delta_F}$ on the y-axis to show how extreme the
protocluster regions are. The protoclusters that stand out the most form the
most massive clusters.
The red line is an empirical fit of $\delta_F \propto -2.9 \log M(z = 0)$.
Bottom: the mass of the most massive halo in the protocluster. The red line
assumes linear growth, $M(z = 0) \propto M(z = 2.5)$. The high mass clusters
appear to grow faster than the linear scaling, although this could be due to
small numbers.}
  \label{fig:mz0}
\end{figure}

In Figure~\ref{fig:mz0}, we show three protocluster properties vs.\ the
resulting cluster mass $M(z = 0)$. We plot the individual protocluster values
with light gray dots, and $M(z = 0)$ binned results (with std.\ dev.\ error
bars) in black. The red dashed lines show approximate scalings for each
quantity. The top panel shows the protocluster half-mass radii, which scales
with the cluster mass. We expect the half-mass and virial radii to scale
similarly with mass. The red line shows the $r \propto M^{1/3}$ relation, which
fits the protocluster sizes well. This falls in line with the
expectation that more massive clusters form from larger overdense regions. The
second panel shows the protocluster flux decrement
$\delta_F / \sigma_{\delta_F}$, evaluated at the protocluster centers from the
$4 \, h^{-1}$Mpc smoothed flux field. In this case, the red line is entirely
empirical. We noticed that the flux decrement scales roughly linearly with
$\log M(z = 0)$ and found a good fit using
$\delta_F \propto -2.9 \log M(z = 0)$. This means that more massive clusters
stand out more significantly in the flux field, although the flux decrement from
low mass clusters is not very significant. Some low mass clusters have
decrements of only 1 or 2 $\sigma$, which are probably too difficult to
distinguish from other background fluctuations.
Clusters with a mass greater than $3 \times 10^{14} \, h^{-1} \Msun$, however,
mostly originate in regions that are greater than $3 \sigma$ flux decrements.
For this reason, we expect to focus on finding more massive protoclusters.
Finally, in the bottom panel, we show the mass of the most massive protocluster
halo. The red line shows the linear scaling $M(z = 0) \propto M(z = 2.5)$,
although the cluster masses appear to grow a bit faster than this. Although
there is significant scatter in this relationship, this confirms that more
massive progenitor halos form the more massive clusters. This is similar to
what \citet{conroy_et_al_2008} found, where halos roughly maintain mass rank
order as they evolve from $z = 2$ to $z = 0$. Altogether, these trends
suggest that finding progenitors of the most massive clusters will be easiest,
because they host the most massive halos, their flux decrement is more
significant, and because the decrement covers a larger volume.
We check if this expectation holds up in Section~\ref{sec:surveys}.

This section contains a basic characterization of protocluster environments, but
it is important to note that our protoclusters have a wide range of sizes,
profiles and overdensities \citep[see also][]{chiang_et_al_2013}. We have
presented a simplified view of protoclusters focused on properties that will
allow us to identify them in flux maps. The full picture of these environments
is probably much more complex, as illustrated by the examples in
Figure~\ref{fig:progen_slices}. Large statistical samples are required to obtain
a representative view of protocluster formation and the impact of the
protocluster environment on galaxy formation and evolution.

\section{Reconstruction Method}
\label{sec:recon}

We have argued that an efficient method for finding protoclusters is to look for
large-scale decrements in the flux field. In this section we discuss how to make
intermediate-resolution maps of the flux field, suitable for protocluster
searches, from observations of closely-separated sightlines.

We use a Wiener filter \citep{wiener_1949, press_et_al_1992} to estimate the 3D
flux field from the noisy observations along multiple sightlines, as
advocated by \citet{caucci_et_al_2008, lee_et_al_2014a}. The Wiener
filter provides the minimum variance, unbiased linear estimator of the field
(under the assumption of a normal distribution) and
can be used to interpolate the data into regions which are not directly sampled,
making it ideal for our purposes\footnote{See \citet{pichon_et_al_2001} for a
more general method than Wiener filtering and \citet{cisewski_et_al_2014} for
a non-parametric method}. We briefly review the derivation of
the Wiener filter, as we use it, in Appendix \ref{app:wf}, where we also
describe our efficient numerical implementation. Collecting all of the
observations of normalized flux into a data vector, $\vd$, which is the sum of a
signal and noise $\vd = \vs + \vn$, the Wiener filter estimate of the signal at
an arbitrary position is $\vsh = \vL \vd$ with $\vL = \vSmp (\vSpp + \vN)^{-1}$.
Here $\vS$ is the assumed signal covariance, where $m$ and $p$ indicate map or
pixel coordinates, and $\vN$ is the noise covariance. The reconstructed map is
thus
\begin{equation}
  \vsh = \vSmp (\vSpp + \vN)^{-1} \vd
  \label{eq:map}
\end{equation}
Following \citet{caucci_et_al_2008}, we model $\vS$ as the product of two
Gaussians for separations along and transverse to the line-of-sight:
\begin{equation}
  S_{ij} = \sigma_F^2 \exp \left[
    -\frac{ (\mathbf{x}_{\perp, i} - \mathbf{x}_{\perp, j})^2 }{2 l_\perp^2}
    -\frac{ (x_{\parallel, i} - x_{\parallel, j})^2 }{2 l_\parallel^2}
  \right]
  \label{eq:s}
\end{equation}
For the noise covariance, we assume that the pixel-to-pixel noise is
independent, so that $N_{ij} = n_i^2 \delta_{ij}$. These assumptions are
approximations, but they are reasonably accurate in the context of the
Ly$\alpha$ forest and the reconstruction is not sensitive to the form assumed
(see tests in Appendix~\ref{app:covar}). Assuming this form for the signal
covariance and that the noise covariance is diagonal provides a huge advantage
computational advantage, as it allows us to never store the matrices directly
and instead compute them as needed. This reduces the space complexity of the
algorithm from $N^2$ to $N$ so that we can still fit large problems on a single
node. We provide more details of our implementation in
Appendix~\ref{app:implementation}.

In this work, we only discuss reconstructing the flux field since it is
sufficient for our application of finding protoclusters. However, we note here
that other authors have considered schemes to reconstruct the matter density
in the context of galaxies as tomographic tracers \citep{willick_2000,
kitaura_et_al_2009, courtois_et_al_2012} and the Ly$\alpha$
forest \citep{kitaura_et_al_2012} and how to account for redshift-space
distortions in the reconstruction.

\section{Protocluster Identification}
\label{sec:pcid}

As shown in Section~\ref{sec:pc}, protoclusters are significant outliers in
density and flux on scales of several Mpc. In this section, we show how we can
exploit this fact to identify protoclusters in the flux maps.

There are many ways we could test for large-scale outliers, but we start with a
simple process of smoothing with a preferred scale and applying a threshold. We
smooth the flux field with a 3D Gaussian filter, typically with a scale
$\sigma = 4 \, h^{-1}$Mpc. Since the protocluster profiles are roughly Gaussian
with a similar scale, this acts much like a matched filter. We tried running
this procedure with the different $\sigma$ values of 2, 4, and $8 \, h^{-1}$Mpc
and found that the $4 \, h^{-1}$Mpc version performs best. Next, we select all
points below some threshold, and group nearby points together. The grouping
process is also simple, where we merge points within $4 \, h^{-1}$Mpc. This
merging process ensures that we do not mistakenly break up low-flux regions and
also that each region has a buffer from other regions. Finally, for each group
of points, we define a protocluster candidate as a $4 \, h^{-1}$Mpc sphere
centered on the minimum flux point in the group. In principle, we could
adjust the choices of the smoothing scale, merging distance, and candidate
radius independently to optimize the candidate selection, but we found it was
not necessary for our purposes, where this simple procedure already performs
well. For a more advanced identification method,
see the optimal filter presented in Appendix~\ref{app:opt_filt}.

\begin{figure}
  \begin{center}
    \resizebox{\columnwidth}{!}{\includegraphics{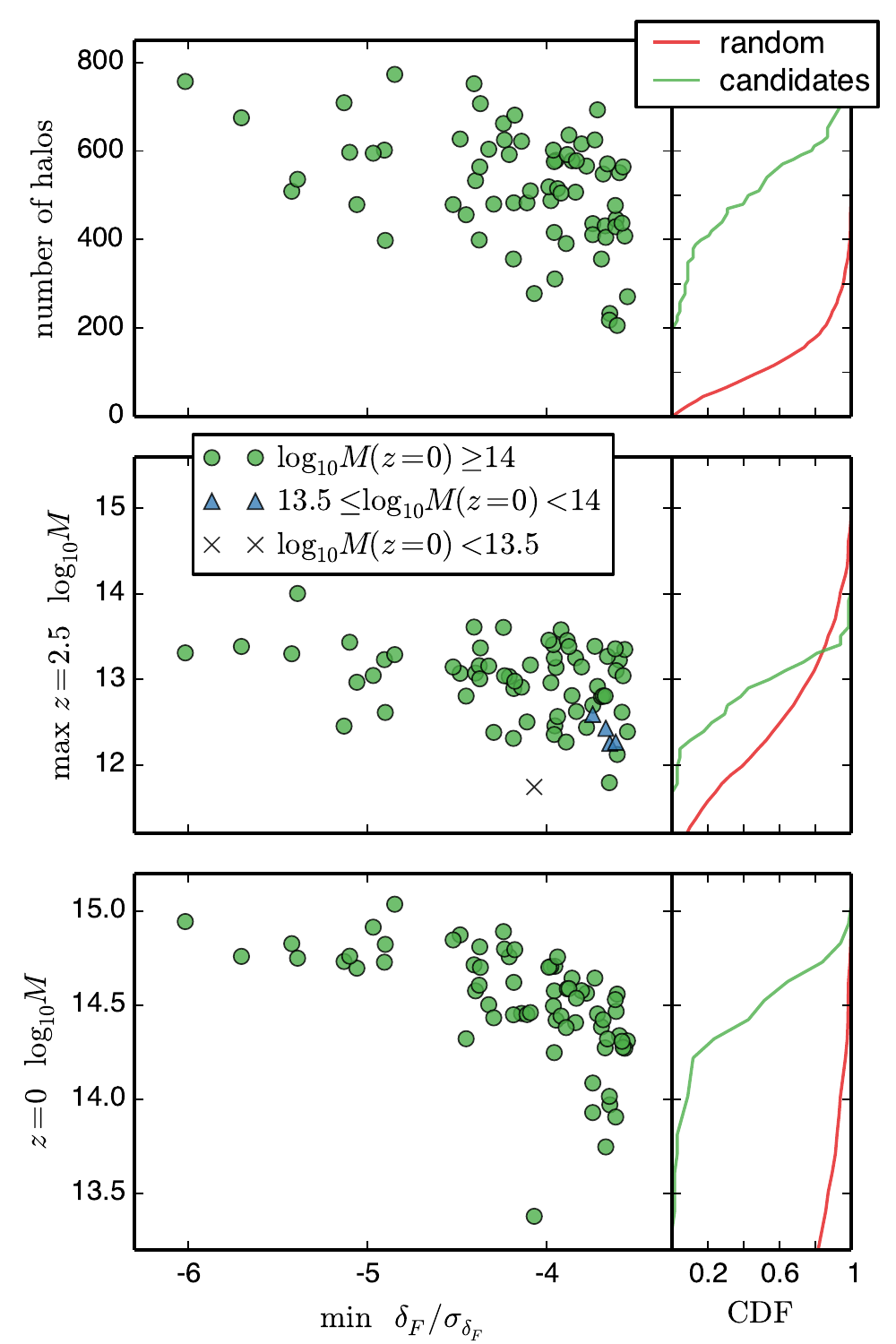}}
  \end{center}
\caption{Halo statistics in $4 \, h^{-1}$Mpc spheres centered on protocluster
candidates (selected by flux decrement) compared to centered on random points.
In each row, we plot the candidates as points on the left and on the right, we
plot the cumulative distribution of the candidates and random points.
Top: The number of halos in the sphere.
Middle: The mass of the maximum mass halo, where the marker indicates the
candidate category based on the $z = 0$ mass (green dots for clusters,
blue triangles for rich groups, and black crosses for anything smaller).
Bottom: The $z = 0$ mass of the maximum mass halo.}
  \label{fig:ideal_halo_stats}
\end{figure}

To get an idea of how this identification procedure performs, we first tested
identifying protocluster candidates from an ideal flux field. We took the
high-resolution flux field from the simulation, smoothed with a $4 \, h^{-1}$Mpc
Gaussian, and downsampled to a typical map resolution (grid spacing) of
$1 \, h^{-1}$Mpc. We chose a threshold of $-3.5$ times the standard deviation of the field, because we found this value performed best for finding protoclusters
forming $> 3 \times 10^{14} \, h^{-1} \Msun$ clusters
(see Figure~\ref{fig:mz0}). When we used more negative threshold values,
we only found the most massive protoclusters, and when we used more positive
threshold values, the protocluster purity decreased and very large protoclusters
were mistakenly merged. These threshold points make up about $10^{-3}$ of the
simulation volume, and were grouped into 68 candidates.

For each candidate, we computed the number of halos in the $4 \, h^{-1}$Mpc
radius sphere and found the maximum mass halo within the sphere. We assigned
each maximum mass halo to a $z = 0$ halo by tracking its particles to $z = 0$
and checking which $z = 0$ halo contained the most of its particles. The
candidate is a protocluster if this $z = 0$ halo mass is $> 10^{14} \, h^{-1}
\Msun$. We also computed these basic halo statistics for randomly positioned
spheres to compare to a field distribution. The results are shown in
Figure~\ref{fig:ideal_halo_stats}, where we plot the candidate number of halos,
maximum $z = 2.5$ halo mass, and resulting $z = 0$ halo mass vs.\ the candidate
$\delta_F / \sigma_{\delta_F}$ value. These panels clearly show that the
low-flux selected candidates have large halo overdensities and are almost all
protoclusters. In the top panel, we plot the number of halos in the protocluster
candidate regions which shows that the candidates are all fairly rich
environments. The candidate regions are on average 5 times the field median
value and the cumulative distributions are well-separated. We note that our
minimum halo mass is about $4 \times 10^{9} \, h^{-1} \Msun$, corresponding to
the requirement that an FoF halo contains at least 50 particles. The middle
panel shows the candidates' maximum-mass halo masses, with markers indicating
the candidate category based on the $z = 0$ mass. The green circles are
clusters, the blue triangles are nearly clusters, and the black cross is a
failure. We also plot the $z = 0$ halo masses in the bottom panel. The four
``nearly'' protocluster candidates have $z = 0$ masses of 5.6, 8.1, 8.5, and
$9.4 \times 10^{13} \, h^{-1} \Msun$, while the failure candidate has a $z = 0$
mass of $2.4 \times 10^{13} \, h^{-1} \Msun$. In all panels, we see the expected
trend that the more significant candidates (in terms of the minimum flux value)
have richer environments and result in more massive clusters. This also
illustrates how the identified protoclusters are more than overdense regions ---
they already host many galaxies and massive galaxies that can be followed up.

The candidate and random sphere cumulative distributions for the maximum $z =
2.5$ mass halos are particularly interesting. Half of the candidate maximum mass
halos are in the mass range of $10^{12}$ -- $10^{13} \, h^{-1} \Msun$ and the
remaining half are in the $10^{13}$ -- $10^{14} \, h^{-1} \Msun$ range. In the
random distribution, half of the maximum mass halos are $< 10^{12} \, h^{-1}
\Msun$, but there is a significant tail to high masses and the distributions
cross at $3 \times 10^{13} \, h^{-1} \Msun$. We checked the total population of
$3 \times 10^{12} \, h^{-1} \Msun$ halos at $z = 2.5$ and found that only 30\%
end up in clusters by $z = 0$. This suggests that our identification procedure
is not just picking out the most massive halos, but finds massive halos with the
right environments to form clusters. This is supported by the cumulative
distributions in the bottom panel, where the candidate and random position
distributions are well-separated again. Despite our simple identification
procedure, these results demonstrate that searching flux maps for large flux
decrements is very effective for finding protoclusters.

We used a fairly conservative threshold value ($-3.5 \sigma_{\delta_F}$) in
order to achieve a high candidate sample purity of 93\%, compared to the random
sample purity of 5\%.
However, this comes with the cost of missing many of the low-mass protoclusters.
We checked the candidate completeness vs.\ cluster mass, and found that above
$3.5 \times 10^{14} \, h^{-1} \Msun$, the completeness is constant and around
80\%. Below this mass, the completeness falls off, reaching 50\% around $2.4
\times 10^{14} \, h^{-1} \Msun$, and 25\% around $1.5 \times 10^{14} \, h^{-1}
\Msun$. As shown in Figure~\ref{fig:mz0}, we expect only more massive
protoclusters (those forming $> 3 \times 10^{14} \, h^{-1} \Msun$ clusters) to
stand out significantly in the flux maps, using a simple threshold at least. At
the same time, the number of clusters quickly increases as we lower the mass,
since these objects are on the tail of the mass function. This unfortunate
combination drives our sample completeness to very small numbers for
protoclusters forming low-mass clusters. However, for moderate-mass
protoclusters (forming $> 3 \times 10^{14} \, h^{-1} \Msun$ clusters), the
method performs well and successfully identifies 70 -- 80\% of the population.

\section{Mock Surveys}
\label{sec:surveys}

In this section, we construct several tomographic mock surveys and run
reconstructions on
the synthetic data to test how our protocluster identification will perform on
realistic data. Specifically, we are interested in what we can achieve with
different values of the average sightline separation $\langle d_\perp \rangle$,
as \citet{lee_et_al_2014a} demonstrated that this is the most important factor
in determining the quality (effective SNR) in the reconstructed maps.
\cite{lee_et_al_2014a} provides a simple relation between the exposure time
$t_{\rm exp}$, the minimum signal-to-noise ratio (SNR) per \AA\
$\mathrm{SNR_{min}}$, and the average sightline separation: $t_{\rm exp} \propto
\mathrm{SNR_{min}}^2 \dperp^{-1.6}$. We assume a fixed $\mathrm{SNR_{min}}$ of
1.5, similar to the recent observations of \citet{lee_et_al_2014c}, so that the
exposure time is just a proxy for the desired average sightline separation. In
principle, we could vary the sightline density and the SNR independently, but in
practice this is not a useful test. If we increase the exposure time to build up
the SNR, it is more advantageous (in terms of the reconstruction quality) to
target fainter sources and increase the sightline density. We initially chose
values of $\dperp = 2$, 2.5, 3, 4, and $6 \, h^{-1}$Mpc. We expect that a
resolution of $2 \, h^{-1}$Mpc will be difficult but possible with existing
instruments, while a spacing of $4 \, h^{-1}$Mpc is fairly coarse, and we
expected $6 \, h^{-1}$Mpc to perform poorly for our application.
We note that the sightlines in \citet{lee_et_al_2014c} have an average
separation of $2.3 \, h^{-1}$Mpc. When we found that the $\dperp = 6 \,
h^{-1}$Mpc separation run still performed decently, we added a survey
configuration meant to mimic the Baryon Oscillation Spectroscopic Survey (BOSS)
survey \citep{dawson_et_al_2013}. For the BOSS-like configuration, we chose an
average sightline separation of $15 \, h^{-1}$Mpc, which is roughly the spacing
for the $200 \, \mathrm{deg}^2$ of the survey with a source density of 1.5 -- 2
times the mean.

We construct mock surveys using our full $(256 \, h^{-1} \mathrm{Mpc})^3$ box.
We first choose skewer positions by drawing random $(x, y)$ coordinates in the
box. We take the ideal $F$ values along the skewer, smooth the signal based on a
typical instrumental resolution $R = 1100$, and bin in pixel widths of $1.2 \,
$\AA. We call this smoothed and binned flux $F_{\rm inst}$. For each spectrum,
we choose a constant per pixel SNR. We draw a random SNR
value from a simple SNR distribution described below. Next, we realize noise for
each spectrum based on its per pixel SNR value. For each pixel, we draw a random
noise value from a normal distribution with scale $\sigma = \mf / \SNR$. We add
the noise vector $F_{\rm noise}$ to $F_{\rm inst}$ to get the final mock fluxes
$F_{\rm syn}$. Altogether, the input to the reconstruction includes the pixel
positions $\vx$, the data vector $\vd = \vec F_{\rm syn} / \mf - 1$, and the
noise vector $\vn = \vec 1 / \SNR$.

We model the sightline SNR distribution as a power law, with a scaling based on
the LBG luminosity function and the observed distribution in
\citet{lee_et_al_2014c}. We define the number of sightlines per deg$^{2}$ as
$n_{\rm los} = (70 \, h^{-1} \mathrm{Mpc} / \dperp)^2 \, \mathrm{deg}^{-2}$ for
our cosmology and $z = 2.5$. Our model is $d n_{\rm los} / d \SNR \propto
\SNR^{-\alpha}$, and we want to determine values of $\alpha$. Based on fits to
the LBG luminosity function, \cite{lee_et_al_2014a} found that $d \log n_{\rm
los} / d g$ is close to unity for the sources we are interested in, where $g$ is
the source g-band magnitude. Combined with the relation $d \log \SNR / d g =
-2.5$, we have $\alpha = -d \log n_{\rm los} / d \log \SNR = 2.5$. This is a
good approximation, but as we probe brighter in the luminosity function and sit
more on the exponential tail, we know that $|d \log n_{\rm los} / d g|$ must
increase. To correct for this, we take the SNR distribution from our pilot
observations, rescale them based on $\SNR_{\rm new} / \SNR_{\rm obs} =
(\dperp_{\rm new} / \dperp_{\rm obs})^{-0.8}$, and fit a power law. For our
choices of $\dperp = 2$, 2.5, 3, and $4 \, h^{-1}$Mpc, we found $\alpha = 2.7,
2.9, 3.5, 3.6$. For larger separations, we did not have enough bright sources in
the pilot observations to reliably estimate $\alpha$, so we kept $\alpha = 3.6$.
We note that for large separations, we would also target more QSOs, which have a
smaller $|d \log n_{\rm los} / d g|$ value at these magnitudes, and provides
a natural maximum value for $\alpha$.

Altogether, we ran 30 mock surveys and reconstructions. For each choice of
$\dperp$, we ran 5 reconstructions to check how the results varied with a fixed
ideal $\delta_F$, but different skewer sampling and noise realizations. For all
reconstructions, we fixed $\sigma_F^2 = 0.05$, $l_\parallel = 2 \, h^{-1}$Mpc,
and $l_\perp = \dperp$ as done in \citet{lee_et_al_2014a}. The small-separation
runs were much more time consuming than the large-separation runs since the
$\Npix \propto \langle d_\perp \rangle^{-2}$ and the algorithm scales with
$\Npix^2$, so that a run with a half the average sightline separation takes 16
times longer.

\begin{table}
  \begin{center}
    \caption{Protocluster candidates and success rates}
    \begin{tabular}{l l r r r r}
      \toprule
      Map & $t_{\rm exp}$ (hrs) & $N_{\rm cand}$ & $f_{\rm PC}$
      & $f_{\rm NPC}$ & $f_{\rm fail}$ \\
      \midrule
      ideal & N/A & 68 & 0.93 & 0.06 & 0.01 \\
      random spheres & N/A & 1000 & 0.05 & 0.07 & 0.88 \\
      \cmidrule(r){1-2}
      $\dperp = 2 \, h^{-1}$Mpc & 2.7 & 73 & 0.89 & 0.08 & 0.03 \\
      $\dperp = 2.5 \, h^{-1}$Mpc & 1.9 & 68 & 0.89 & 0.09 & 0.01 \\
      $\dperp = 3 \, h^{-1}$Mpc & 1.4 & 76 & 0.84 & 0.10 & 0.06 \\
      $\dperp = 4 \, h^{-1}$Mpc & 0.90 & 77 & 0.78 & 0.15 & 0.07 \\
      $\dperp = 6 \, h^{-1}$Mpc & 0.47 & 72 & 0.61 & 0.20 & 0.20 \\
      $\dperp = 15 \, h^{-1}$Mpc & N/A & 26 & 0.35 & 0.10 & 0.55 \\
      \bottomrule
      \label{tab:cand_stats}
    \end{tabular}
  \end{center}
Protocluster identification success rates for the ideal $\delta_F$ field and
randomly-positioned spheres compared to the mock survey reconstructions.
$t_{\rm exp}$ is the corresponding exposure time to achieve the desired
sightline spacing (rescaled from the Keck/LRIS setup in \citet{lee_et_al_2014c}).
$N_{\rm cand}$ is the number of candidates found in the map and the $f$ values
are the fractions of candidates broken into three class: protoclusters
(PC), nearly protoclusters (NPC), and failures (fail). The numbers reported for
the mock reconstructions are averages over the 5 realizations of sightline
positions and noise. The $\dperp = 15 \, h^{-1}$Mpc configuration is meant to
reproduce the relatively high sightline density areas of the BOSS survey.
\end{table}

We tested the success of the surveys by running the protocluster identification
procedure on the mock maps and comparing to the halo catalog, just as we did for
the ideal field in the previous section. Again, we used a smoothing scale of $4
\, h^{-1}$Mpc, a threshold of $-3.5$ times the standard deviation, and a region
size of $4 \, h^{-1}$Mpc. Overall, we found an good agreement between
protocluster candidates in the ideal and reconstructed fields, and that the
success rates decrease with increasing average sightline spacing, as expected.
In Table~\ref{tab:cand_stats}, we list the number of candidates identified in
each map, and the fraction of candidates that fell into classes of protoclusters
(PC), nearly protoclusters (NPC), and failures (fail). These classes follow the
definitions used earlier in Figure~\ref{fig:ideal_halo_stats}, where
protoclusters form clusters ($M \ge 10^{14} \, h^{-1} \Msun$), nearly
protoclusters almost form clusters ($10^{13.5} \, h^{-1} \Msun \le M < 10^{14}
\, h^{-1} \Msun$), and failures are anything less massive ($M < 10^{13.5} \,
h^{-1} \Msun$). The mock results are averaged over the 5 survey realizations for
each configuration. The number of candidates in the reconstructed maps is
consistent with the result for the ideal field, although slightly higher, except
for the BOSS-like survey which is much lower. If we scale the number of
candidates ($N_{\rm cand} \sim 70$) found in the simulation volume of $(256 \,
h^{-1} \mathrm{Mpc})^3$ to the final CLAMATO volume of $70 \times 70 \times 230
\, (h^{-1} \mathrm{Mpc})^3$, we should find 5 candidates. However, using a
smaller threshold will yield many more candidates, if the decrease in purity can
be accommodated.

There is a clear trend of the success rates vs.\ the average sightline
separation. As the sightline separation increases, the map quality decreases,
and the sightlines begin to miss protocluster structures leading to the decline
in success. Additionally, as the noise in the map increases, the false positive
rate increases. When we increase the sightline separation to larger than $10 \,
h^{-1}$Mpc, the quality of the map degrades significantly, which is reflected in
the BOSS-like success rates and lower number of candidates. For small
separations, the protocluster identification success rate is close to ideal ---
93\% in the ideal case and 89\% for $\dperp = 2$ and $2.5 \, h^{-1}$Mpc. Even
with a coarse sightline separation of $4 \, h^{-1}$Mpc, the success rate is
78\%, and this only drops to 60\% with the $6 \, h^{-1}$Mpc separation that we
thought might be catastrophic.

In the BOSS-like separation surveys, the candidate purity is much lower. This is
expected since the average spacing in this case is larger than all but the
largest protoclusters. However, with random positions, it is possible for
several sightlines to overlap with a protocluster and this configuration still
performs significantly better than random. We believe the purity in the
BOSS-like configuration could also be improved if we considered sightline
positions, and only saved candidates with many overlapping sightlines.

\begin{figure}
  \begin{center}
    \resizebox{\columnwidth}{!}{\includegraphics{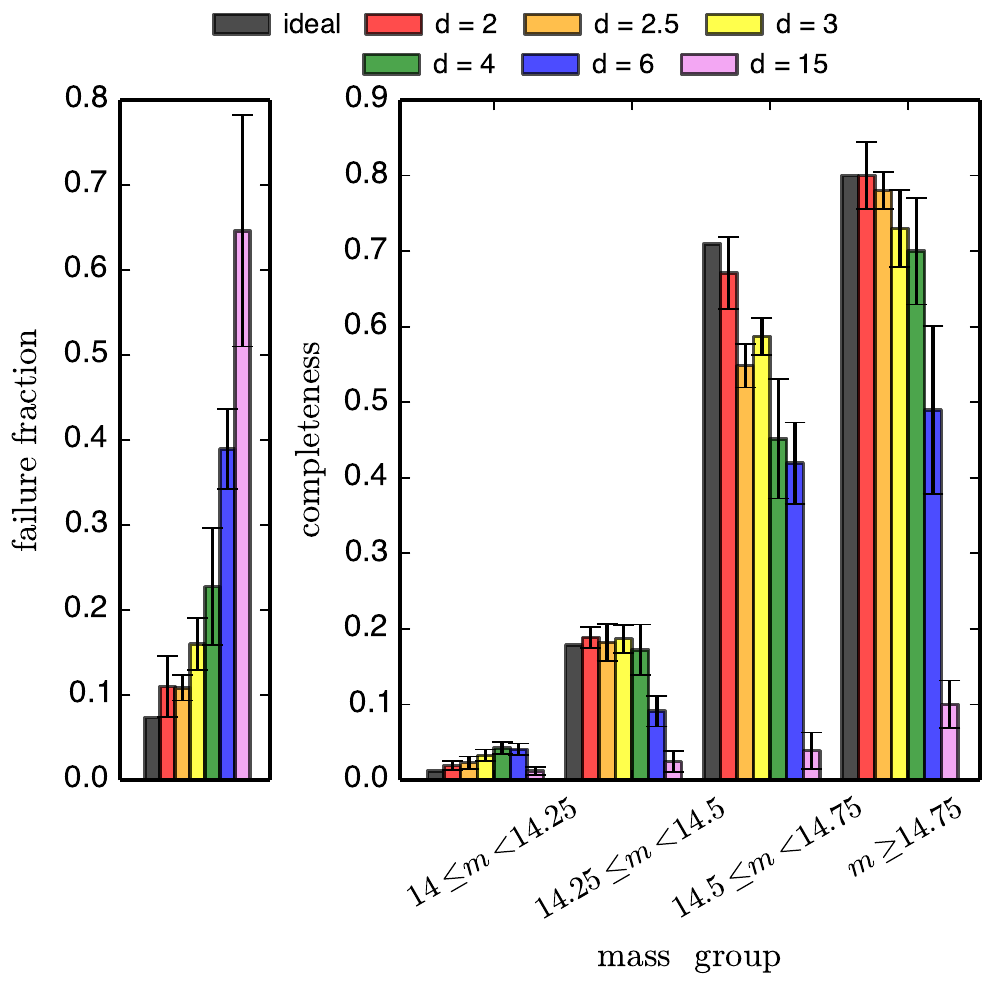}}
  \end{center}
\caption{Candidate failure fractions and completeness identified in the ideal
and mock maps.
Left: Fraction of candidates that are failures ($M(z = 0) < 10^{14} \, h^{-1}
\Msun$).
Right: Candidate completeness measured in 4 cluster mass bins, where the mass
range in $m = \log_{10}[M / (h^{-1} \Msun)]$ is indicated on the x-axis.
The mock map counts are averages over 5 realizations, with std.\ dev.\
error bars. As the sightline separation increases, we see a steady increase in
contamination, and the success fraction decreases. In the low-mass cluster
bins, the mock survey completeness is sometimes higher than the noiseless map.
This is a result of noise in the mock reconstructions pushing some less
significant protoclusters over the chosen threshold value.}
  \label{fig:cand_mass}
\end{figure}

In Figure~\ref{fig:cand_mass}, we show the candidate completeness and failure
rates for the various survey configurations. On the left, we plot the fraction
of candidates that did not form clusters. The mock map values are averages over
the 5 realizations and we show the std.\ dev.\ error bars. We see a steady
increase
in the candidate contamination as the sightline separation increases. On the
right, we plot the candidate completeness measured in four cluster mass bins.
For reference, the numbers of protoclusters from the full sample in these bins
are 251, 123, 31, and 20. In the two high mass bins, the completeness of the
$\dperp = 2$, 2.5, and $3 \, h^{-1}$Mpc surveys is similar to the ideal result.
The completeness decreases for larger separations, although it is still about
50\% for the $\dperp = 4$ and $6 \, h^{-1}$Mpc, but only 5\% -- 10\% for the
BOSS-like survey. For the two low mass bin, the completeness overall is much
lower, as discussed in the previous section. The completeness falls off for very
large separations, as before, but for small separations, the completeness is
sometimes larger than the ideal map. This is due to the noise in the
reconstructions scattering some low mass protoclusters over the threshold value.
That is, for some protoclusters that did not make the cut in the noiseless map,
the reconstruction noise fortunately pushes them over the edge. Overall, this
result makes us confident that we can find a large fraction of the protoclusters
that form $> 3 \times 10^{14} \, h^{-1} \Msun$ clusters with a CLAMATO-like
survey.

\begin{figure*}
  \begin{center}
    \resizebox{\textwidth}{!}{\includegraphics{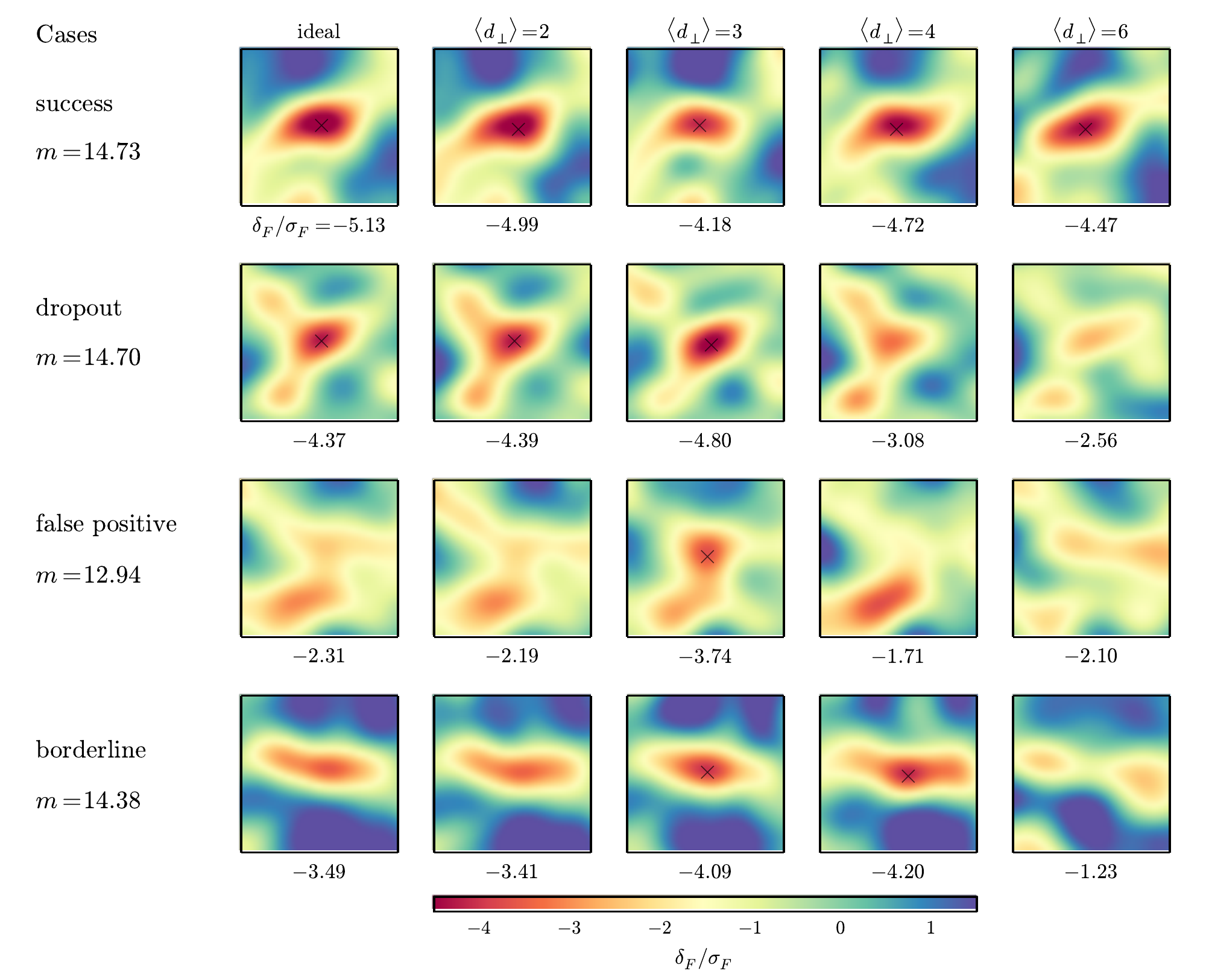}}
  \end{center}
\caption{Slices from the ideal and mock maps centered on four candidates
scenarios. The images show the $\delta_F / \sigma_{\delta_F}$ values in slices
that are $40 \, h^{-1}$Mpc across and $2 \, h^{-1}$Mpc thick.
From left to right, we show the ideal field and the mock reconstructions with
increasing $\dperp$, and we mark successful identifications with a black cross.
We also annotate the $\delta_F / \sigma_{\delta_F}$ value in the center of the
map under each image. For each row, we annotate the case name and candidate
mass $m = \log_{10} [M / (h^{-1} \Msun)]$ on the left.
Top row: A success case where the protocluster is identified in all maps.
Second row: As the average sightline separation increases, the sightlines do
not sample the low flux region well enough and the candidate ``drops out''.
Third row: A moderately low flux region with additional noise can create a
false positive.
Bottom row: A borderline protocluster where the reconstruction noise scatters
the candidate over the threshold value.
There is another failure case (bad merge) not shown here, but explained in the
text.}
  \label{fig:cand_cases}
\end{figure*}

In order to understand the cases where our protocluster identification method
failed (either missing protoclusters or selecting false positives), we looked at
many slices of individual candidates. We performed a union of all candidates
identified in the ideal map and in the mock reconstructions, based on the
candidate's $z = 0$ halo ID, and tracked which candidates were identified in
which maps. After visually inspecting many candidates, we found that we could
group the failures into four categories which we called dropout, bad merge,
false positive, and borderline protocluster. We illustrate these cases with
example candidates in Figure~\ref{fig:cand_cases}. Each row is a separate
candidate, and the columns show the same slice from the ideal map and the
$\dperp = 2$, 3, 4, and $6 \, h^{-1}$Mpc mock maps. If the candidate was
identified in the map, we marked the center with a black cross. We also
annotated the $\delta_F / \sigma_{\delta_F}$ value from each map (at the
candidate center) under the image. In the top row, we show a successful case,
where the candidate forms a massive cluster, and the protocluster is found in
all of the maps. This case was not very common when we included the large $6 \,
h^{-1}$Mpc separation maps, but it was usually the case for the most massive
protoclusters that created a significant ($> 5 \sigma$) flux decrement.

The first failure case, dropout, is the most common scenario for a missed
protocluster identification. The protocluster creates a clear flux decrement in
the ideal map and small separation survey maps, but the signal drops out in the
large separation survey maps. An example is shown in the second row of
Figure~\ref{fig:cand_cases}. In the example shown, the protocluster is
successfully identified in the ideal and small $\dperp$ maps, but as the
sightline separation increases, the region is less well-sampled and the flux
values in the region never drop below the threshold. We also found plenty of
cases where the candidate is missed in the $\dperp = 3$ or $4 \, h^{-1}$Mpc
maps, but found again in the larger separation maps, just due to how the
sightlines and the protocluster line up in a given random survey realization.

The second failure case, bad merge, is another scenario that results in missing
a protocluster, and is due to a weakness in our method for merging points during
the identification procedure. We found a few cases where two protoclusters were
linked by a dense filament, so that the two regions that should have been
separate candidates were mistakenly merged. The grouped points were usually
similar shapes in the different maps, but the flux minimum could end up in
either protocluster depending on the reconstruction noise. If these candidates
were correctly partitioned during the merging step, there would be another
successful identification in each map.

In regions of moderately low flux, it is possible for the reconstruction noise
to scatter low, and create false positives. This scenario is origin of the
increasing contamination (failure fraction) with increasing sightline
separation. The third row of Figure~\ref{fig:cand_cases} shows an example of
this, where the candidate is not a protocluster, but is mistakenly identified in
one of the maps. Of course, sometimes this reconstruction noise also scatters
low mass protoclusters in the right direction. This is the origin of the final
scenario we called borderline protoclusters. In this case, the protocluster
creates a flux decrement just under the threshold, so that it is not identified
in the noiseless map, but noise can scatter it over the threshold, so that it
is identified in the mock reconstruction. An example is shown in the bottom row
of Figure~\ref{fig:cand_cases}, where the flux decrement in the noiseless map is
just under the threshold. In the $\dperp = 3$ and $4 \, h^{-1}$Mpc
reconstructions, the noise scatters the flux decrement to over $4 \sigma$, which
makes it a successful candidate.

\section{Conclusions}
\label{sec:conclusions}

In this paper, we characterized the signature of protoclusters at $z \simeq 2$
-- 3, and demonstrated the success of a simple method for finding these
protoclusters from the associated Ly$\alpha$ forest flux decrement. The
tomographic reconstruction of the 3D Ly$\alpha$ forest transmitted flux field
from individual sightlines is the crucial step to this method. In order to
handle datasets with large numbers of pixels, we implemented a new fast Wiener
Filter code, which we are making publicly available. This code will make it
possible to run reconstructions on the scale of $\Npix \gtrsim 10^6$, larger
than the expected size of the ongoing CLAMATO survey.

We identified protoclusters at $z = 2.5$ using a large cosmological $N$-body
simulation with sufficient resolution to capture individual absorption systems
comprising the Ly$\alpha$ forest and covering enough volume to contain a
respectable cluster sample. We constructed FoF halo catalogs for each simulation
snapshot and defined clusters at $z = 0$ with a mass cut of $M \ge 10^{14} \,
h^{-1} \Msun$. We then identified protoclusters by tracking cluster member
particles from $z = 0$ back to $z = 2.5$ (by particle ID) and characterized the
protocluster regions. The key signature of protoclusters is that they are
outliers in density and flux on large scales. We found that protocluster centers
are above the 95$^{\rm th}$ percentile of the density and flux decrement and
that the half-mass radius of typical protoclusters at this redshift is $4 \,
h^{-1}$Mpc. The density and flux profiles of protocluster regions are well fit
by a Gaussian with a scale of $5 \, h^{-1}$Mpc, suggesting that maps with
several Mpc resolution should easily resolve these structures. We also found
that the flux decrement and radius of a protocluster increases with its $z = 0$
mass, so that it is easiest to find the protoclusters that form the more massive
clusters.

We reviewed our tomographic reconstruction method (a Wiener Filter) and some
specifics of our application. Specifically, we assume a certain form of the
signal covariance and that the noise covariance is diagonal. These assumptions
significantly reduce the space complexity of our algorithm, so that we can
easily fit the calculations on a single node, avoiding significant communication
costs and taking advantage of shared-memory parallelism. Additionally, this
design will easily take advantage of upcoming compute architectures, where the
number of cores per node is expected to increase in the near future, and could
easily be extended to run on GPUs.

We designed a procedure to identify protocluster candidates in the flux maps. To
choose candidates we smooth the map, apply a threshold, and group the remaining
points into candidates. We ran the procedure on noiseless maps and compared to
the simulation halo catalog, finding that we can achieve 90\% candidate purity
with this simple method. We also confirmed that the method tends to find
protoclusters that form the most massive clusters ($> 3 \times 10^{14} \, h^{-1}
\Msun$). The most massive halos in the identified protoclusters have masses of
about $10^{13} \, h^{-1} \Msun$ --- still very difficult to find at these
redshifts using alternative methods.

Finally, we created realistic mock surveys (similar to the recent observations
of \citet{lee_et_al_2014c}) and reconstructed the flux maps with our code. We
found that surveys with an average sightline spacing $\dperp = 2.5 \, h^{-1}$Mpc
performs essentially the same as the ideal, noiseless map. Such surveys should
identify protoclusters with a 90\% success rate, and find 70 -- 80\% of the
protoclusters that form clusters with masses $> 3 \times 10^{14} \, h^{-1}
\Msun$). Using the same conservative threshold, we would identify 5
protoclusters in the planned CLAMATO volume. However, the volume should contain
about 30 protoclusters including those that form lower mass clusters.

Finding protoclusters at $z \simeq 2$ -- 3 remains an observationally
challenging problem. With relatively simple methods, we have demonstrated a
promising new technique for finding protoclusters at these redshifts. As shown
in \citet{lee_et_al_2014a} IGM tomography offers a novel method for mapping
large volumes with high efficiency using existing facilities. The method can
return large samples of protoclusters and does not suffer from projection
effects (or redshift errors). The Ly$\alpha$ forest also has the advantage of
only probing mildly nonlinear densities, allowing for \textit{ab initio}
calculation of the density-observable relation (i.e.\ the bias) via numerical
simulations. Future work can easily extend this to reconstruct density maps,
include redshift-space distortions, and incorporate more advanced models of
protoclusters.

We thank Andreu Font-Ribera and Zarija Luki\'{c} for useful discussions. The
simulation, mock surveys, and reconstructions discussed in this work were
performed on the Edison Cray XC30 system at the National Energy Research
Scientific Computing Center, a DOE Office of Science User Facility supported by
the Office of Science of the U.S. Department of Energy under Contract No.
DE-AC02-05CH11231. This research has made use of NASA's Astrophysics Data System
and of the astro-ph preprint archive at arXiv.org.

\bibliographystyle{apj}
\bibliography{ms}

\appendix

\section{Reconstruction Derivation and Implementation}

In this appendix section we briefly review the Wiener filter, to establish our
notation, and describe our efficient numerical algorithm for map making.

\subsection{Wiener filter}
\label{app:wf}

We assume our data is made up of the signal we are interested in and additive
noise $\vd = \vs_p + \vn$. In order to keep coordinates clear, we use a $p$
subscript to indicate `pixel' coordinates, and an $m$ subscript to indicate
`map' coordinates. Note that some other texts characterize this difference with the
instrumental response matrix $\vec R$ as $\vs_p = \vec R \vs_m$. We want to make
a linear estimate of the signal $\vsh = \vL \vd$, with minimal error $\epsilon =
E [|\vs_m - \vsh|^2]$. We start by simplifying the error expression.
\[
  \epsilon = \tr \left( E[\vs_m \vs_m^T] - E[\vs_m \vsh^T] - E[\vsh \vs_m^T]
    + E[\vsh \vsh^T] \right)
\]
The first term $E[\vs_m \vs_m^T]$ is just the signal covariance $\vSmm$. The
second term is
\[
  \begin{aligned}
    E[\vs_m \vsh^T] &= E[\vs_m (\vL \vd)^T] = E[\vs_m \vd^T \vL^T]
                   = E[\vs_m (\vs_p^T + \vn^T) \vL^T] \\
                  &= (E[\vs_m \vs_p^T] + E[\vs_m \vn^T]) \vL^T = \vSmp \vL^T
  \end{aligned}
\]
since we assume $E[\vs_m \vn^T] = 0$. By a similar manipulation the third term
$E[\vsh \vs_m^T] = \vL \vSpm$. The fourth term is
\[
  \begin{aligned}
    E[\vsh \vsh^T] &= \vL E[\vd \vd^T] \vL^T \\
      &= \vL (E[\vs_p \vs_p^T] + E[\vs_p \vn^T] + E[\vn \vs_p^T] + E[\vn \vn^T]) \vL^T \\
      &= \vL (\vSpp + \vN) \vL^T
  \end{aligned}
\]
Altogether, the error is
\[
  \begin{aligned}
    \epsilon &= \tr \vSmm - \tr(\vSpm \vL^T) - \tr(\vL \vSmp)
         + \tr(\vL (\vSpp + \vN) \vL^T) \\
      &= \tr \vSmm - 2 \tr(\vL \vSmp) + \tr(\vL (\vSpp + \vN) \vL^T)
  \end{aligned}
\]
Taking the derivative of the error with respect to the operator, we have
\[
  \frac{\partial \epsilon}{\partial \vL} = - 2 \vSmp + 2 (\vSpp + \vN)^T \vL^T
\]
And then evaluating $\partial \epsilon / \partial \vL = 0$ to find the minimum
error, we have the optimal operator $\vL = \vSmp (\vSpp + \vN)^{-1}$.

\subsection{Signal covariance}
\label{app:covar}

The form we assume for the signal covariance is a product of two Gaussians, as
shown in Equation~\ref{eq:s}. The flux correlation function should roughly have
this form, but it is certainly not correct in detail. In this section, we
consider the difference between the true signal covariance and our model (with
appropriate $l_\perp$ and $l_\parallel$ values), and how this model inadequacy
might affect our reconstruction results.

\begin{figure*}
  \begin{center}
    \resizebox{\textwidth}{!}{\includegraphics{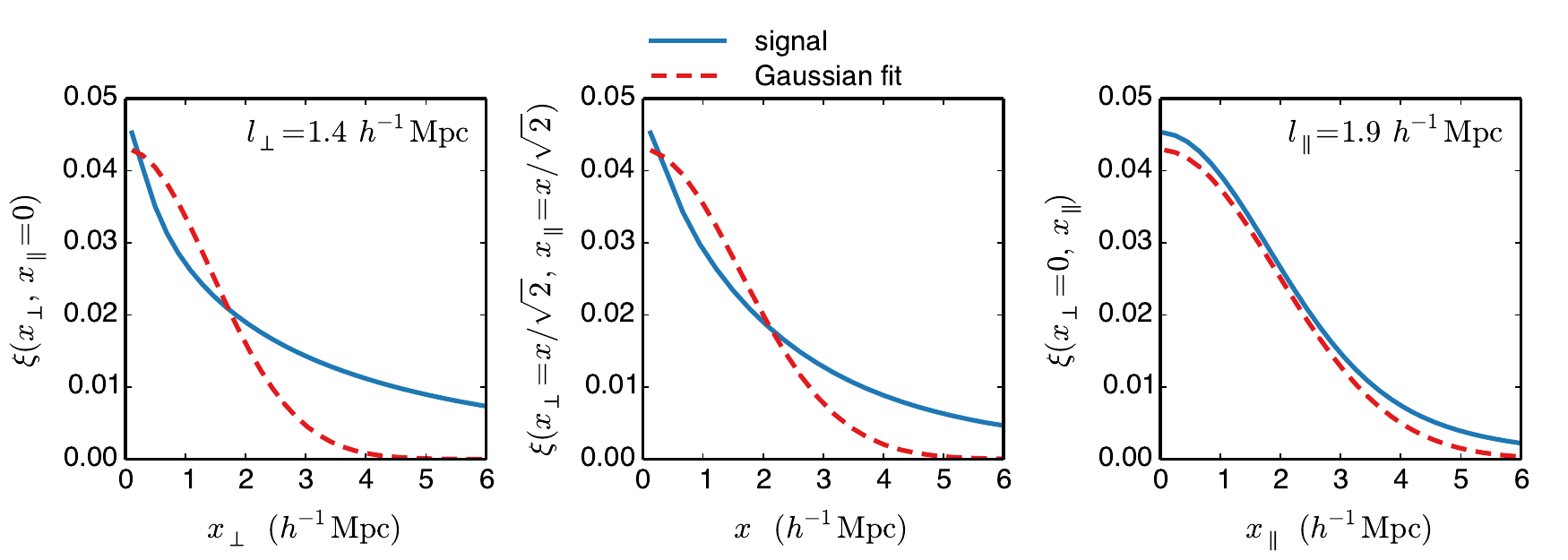}}
  \end{center}
\caption{The signal correlation function (solid line) compared to our assumed
Gaussian form (dashed line), with amplitude, $l_\perp$, and $l_\parallel$ fit
to the signal data.}
  \label{fig:flux_corr}
\end{figure*}

In Figure~\ref{fig:flux_corr}, we compare the correlation function of $\delta_F$
from the simulation (labeled signal) and our model fit to the ideal signal
(labeled Gaussian fit). We have smoothed the signal along the line of sight to
match a typical spectrograph resolution ($R \sim 1100$). From left to right, the
three panels show different slices through the $(x_\perp, x_\parallel)$ plane,
first all perpendicular, for $x_\perp = x_\parallel$, and for all parallel. We
also annotated the fit Gaussian scales $l_\perp$ and $l_\parallel$. The Gaussian
product shape does well along the line of sight, due to the fact that we have
mocked the instrumental smoothing with a Gaussian filter, and the unsmoothed
flux correlation is small for scales larger than the filter scale. Across the
line of sight, our model does much worse. In a future iteration, we will
consider using a sum of Gaussians...
Such a mismatch between the simulation and model might be worrying,
but we argue that this is not a concern for our application. In the case of
Wiener filtering, most elements of the operator
$\vS (\vS + \vN)^{-1}$ are close to 0 or 1, and the shape of $\vS$ only changes
values in the intermediate regime \citep[see e.g.][for discussion]{press_et_al_1992}.

\begin{figure}
  \begin{center}
    \resizebox{4in}{!}{\includegraphics{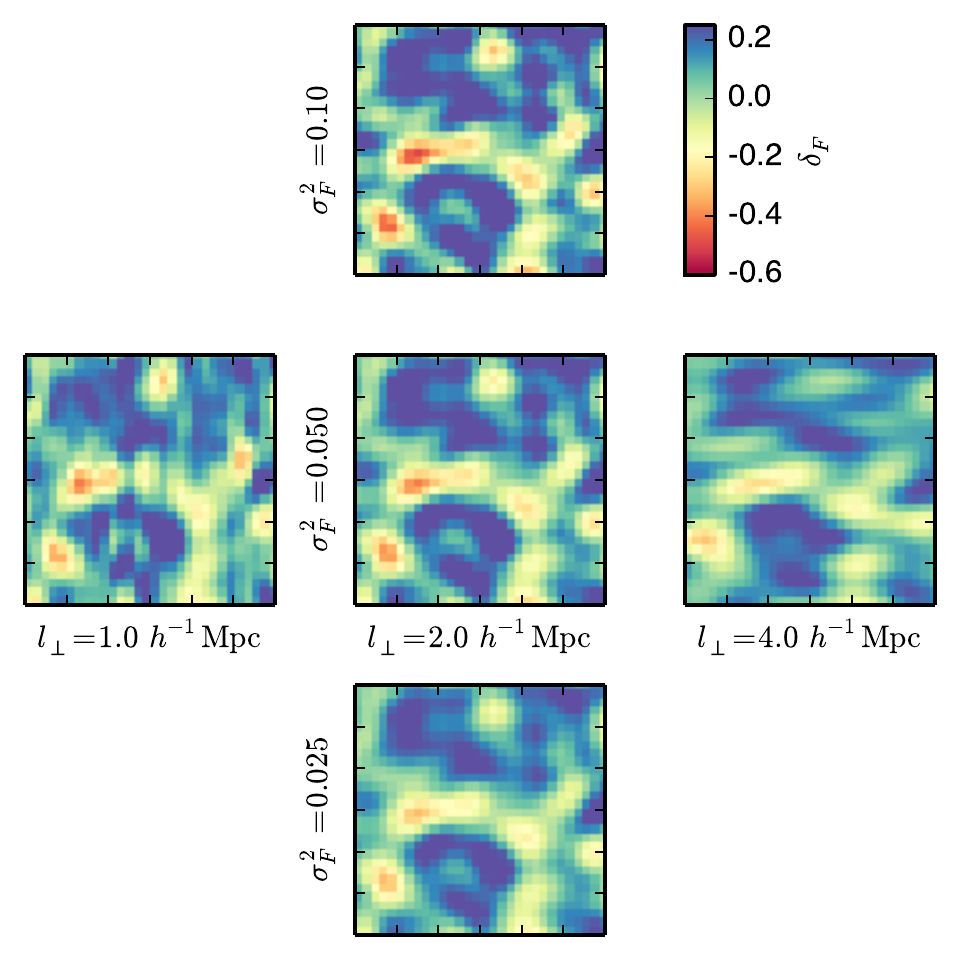}}
  \end{center}
\caption{Slices through reconstructions run on the same pixel data, varying the
signal covariance parameters $l_\perp$ and $\sigma_F^2$. The slices are $2 \,
h^{-1}$Mpc thick, projected over the x-axis (perpendicular to line of sight). The
vertical axis of the images is along the line of sight. From left to right, we
double the $l_\perp$ value, from 1 to 4. From bottom to top, we double the
$\sigma_F^2$ value, from 0.025 to 0.1. Any reasonable deviations from the
best case parameter values do not affect the morphology of the resulting map.}
  \label{fig:sc_slices}
\end{figure}

In order to test the effect of an inaccurate covariance assumption on the
reconstruction, we ran several reconstructions of the same pixel data, varying
the signal covariance parameters $l_\perp$ and $\sigma_F^2$. The same slice from
each reconstruction is shown in Figure~\ref{fig:sc_slices}. We vary $l_\perp$
from left to right and $\sigma_F^2$ from top to bottom. Overall, it appears
that any reasonable changes to the parameter values (relative to the best-fit
values) do not affect the morphology of the structures in the map. Increasing
the flux variance increases the variance in the final map. This is due to the
increase in all $\vS$ elements relative to $\vN$ so that pixels have larger
weights in the reconstruction. That is, increasing the flux variance parameter
should have the same effect as reducing all pixel noise estimates. Varying
the correlation scale $l_\perp$ (or $l_\parallel$) has a more dramatic effect.
With a fixed sightline sampling and a smaller correlation scale, the noise will
obviously have a larger effect on the map, as the pixels are less correlated.
As we increase the correlation scale, structures become increasing smoothed out.
We found that changes in $l_\parallel$ behave the same as changes in $l_\perp$,
so we did not add it to the plot.

\subsection{Numerical Implementation and Scaling}
\label{app:implementation}

Computationally, the map making process consists of two steps. First, there is
the matrix inversion and matrix-vector multiply $\vx = \vA^{-1} \vb = (\vSpp +
\vN)^{-1} \vd$. The second step of the map process is just the multiplication
$\vm = \vSmp \vx$. The matrix $\vA$ is symmetric and positive definite, so there
are several computationally efficient methods for obtaining the solution $\vx$.
Since our signal and noise matrices are both relatively sparse, we use the
preconditioned conjugate gradient (PCG) method with a Jacobi pre-conditioner
\citep{saad_2003}
\footnote{Also see
\url{http://www.cs.cmu.edu/~quake-papers/painless-conjugate-gradient.pdf}}.
PCG is an iterative method which converges rapidly for sufficiently sparse
matrices. For reasonable survey strategies, we do not expect a large number of
pixels within a flux correlation scale (several Mpc), so methods that perform
better for sparse matrices should be advantageous. We use the stopping condition
that the residual is smaller than the norm of the data times a tolerance
parameter, $|\vec r| = |\vb - \vA \vx| < \mathrm{tol} |\vb|$.

The real advantage of PCG for our problem, however, is that it never uses
$\vA$ directly, but only products of $\vA$ and a vector. Since we know the
functional form of $\vSpp$, and we assume $\vN$ is diagonal, we do not have
store the matrix $\vA$, and instead compute elements when needed. This changes
the space complexity of the algorithm from $\Npix^2$ to $\Npix$. For a typical
problem where $\Npix = 10^6$ the difference in storage is about $8\,$TB (for
$\vA$ stored in double precision), demanding several hundred nodes on modern
systems, versus six vectors of length $\Npix$, requiring about $50\,$MB and
easily fitting on a single node. Clearly, the performance of the PCG solve
depends on how quickly we can compute elements of $\vA$. We speed up the element
lookup by using a small table of $\exp(x)$ for the Gaussian. This reduces each
element lookup to 10 add/multiply operations.

Altogether, the cost of the reconstruction algorithm is $N_{\rm lookup} (N_{\rm
iter} \Npix^2  + \Nmap \Npix)$, where $N_{\rm lookup}$ is the number of
operations involved in computing elements of $\vA$ and $N_{\rm iter}$ is the
number of iterations before the PCG reaches the stop condition. We expect
problem sizes of up to $10^6$ pixels and $10^6$ map points, so assuming 100
iterations, the calculation takes $10^{15}$ operations. This estimate indicates
that we will likely not need to parallelize the code beyond shared memory,
especially since the number of cores per node is expected to increase in coming
years.

In order to choose a tolerance value for the PCG stop condition, we tested the
PCG result against a direct Cholesky factorization for small
problems. We generated a mock dataset with $\Npix = 4000$, fixed the pixel
positions and signal, and generated 10 noise realizations with SNR = 5. With
multiple noise realizations, we can estimate the map variance due to noise
compared to the error of the PCG solve. For each of the 10 noise realizations,
we ran the reconstruction with the Cholesky solve and with the PCG solve with
tol values of 1, 0.1, 0.01, and $10^{-3}$. The Cholesky reconstruction took 18
seconds on average while the lowest tol PCG reconstruction took 0.9 seconds on
average. The average number of PCG iterations to reach the various tol values
were 5, 13, 29, and 37. We computed the standard deviation of the Cholesky map
values over noise realizations $\vec{\sigma}_m$ to have a measure of the
variance due to the noise at each map point. The average $\sigma_m$ is 0.06 and
the max is 0.11. We then computed the absolute difference of the PCG maps and
Cholesky maps relative to the map noise std., $\vec \epsilon = |\vsh_{\rm PCG} -
\vsh_{\rm Chol}| / \vec{\sigma}_m$. This error captures the fact that the PCG
error must be smaller for map points with small noise variance. We found that
the errors have an exponential distribution, with maximum values over all map
points of 30, 4.5, 0.47, and 0.059 respectively for the 4 PCG tolerance
settings. Since the max error of the tol = 0.01 PCG maps is less than unity and
the error distribution is exponential, this tolerance setting is in the safe
regime where the PCG residual error in the map is significantly smaller than the
noise. In practice the PCG tol value should be adjusted for the problem at hand
(if the SNR is very different), but this is a conservative choice for Ly$\alpha$
forest data in the near future.

One practical issue with the expressions in Equation~\ref{eq:map} is that it
does not easily allow for masking bad pixels. If we have any pixels with $n =
\mathrm{inf}$, the PCG routine will return nan's. The ability to mask data
is critical for Ly$\alpha$ forest data, where we may run into sky lines that add
significant noise, or any pixels that should be masked entirely. We can rewrite
the map expression using the fact that the noise covariance may be formed as a
product of a lower triangular matrix with its transpose.
In the case of our noise covariance, this $N_{ij} = (n_i \delta_{ik})
(n_j \delta_{jk})^{T}$. It follows that
\begin{equation}
  \vm = \vSmp \vw (\vw \vSpp \vw + \vI)^{-1} \vw \vd
  \label{eq:wmap}
\end{equation}
where $\vw = \vn^{-1}$. In this new expression, the matrix to be inverted is
definite even for pixels with $w = 0$, and the PCG solves will work as expected.
This expression requires more operations than the simpler Equation~\ref{eq:map},
but they add negligible overhead.

The reconstruction code implemented for this work consists of a static library
and a few executables, written in C++, with no dependencies. The code can be
compiled and run with no parallelism, but we recommend enabling OpenMP if
available. The code is publicly available at
\url{http://github.com/caseywstark/dachshund}, and includes some documentation
and a test suite.

\begin{figure}
  \begin{center}
    \resizebox{7.5in}{!}{\includegraphics{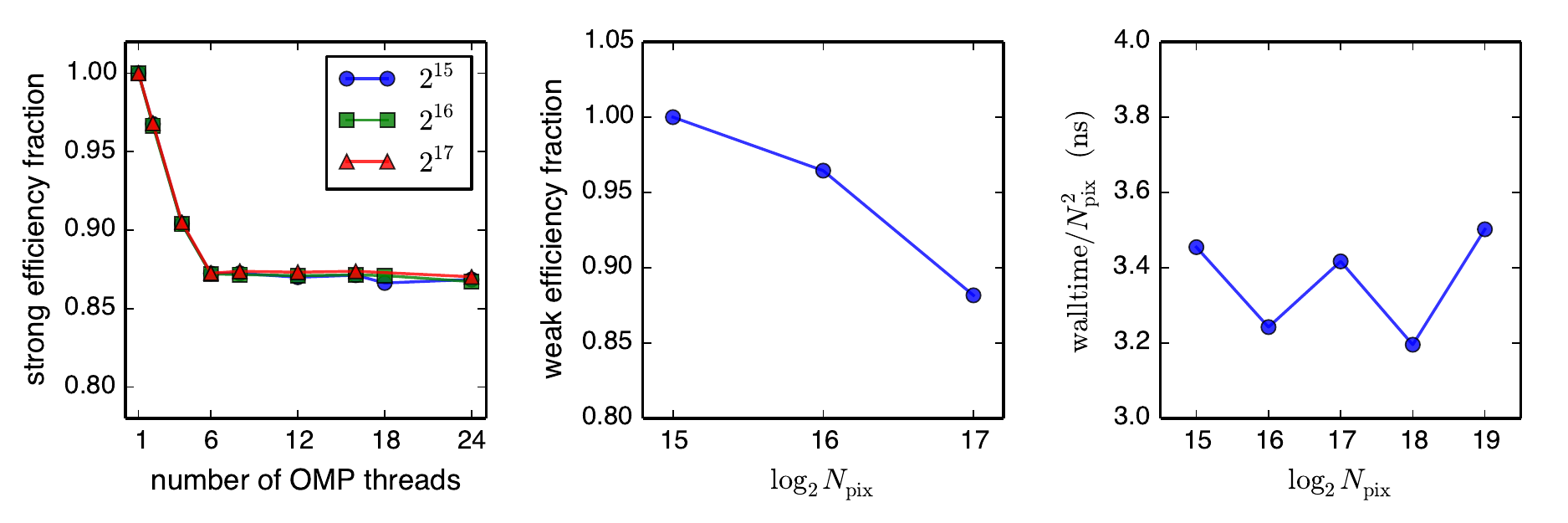}}
  \end{center}
  \caption{Three scaling tests for our code.
  Left: 3 strong scaling tests, where we fix the problem size and increase the
  number of threads. As long as all threads are given enough work, the speedup
  should be linear. We show the efficiency $e_{\rm strong} = t_n / (n t_1)$
  where $n$ is the number of threads and the $t$ is the walltime.
  Middle: a weak scaling test, where we fix the work per thread, increasing the
  problem size and number of threads. We show the efficiency
  $e_{\rm weak} = t_n / t_1$.
  Right: The element-wise time for several problem sizes.}
  \label{fig:scaling}
\end{figure}

We performed scaling tests of our code to give an idea of what problem scale the
code is able to handle within a reasonable wall time. We ran the test problems
on the Edison machine at NERSC. Each Edison node has two 12-core Intel ``Ivy
Bridge'' processors clocked at 2.4 GHz. We created mock surveys like  the ones
in Section~\ref{sec:surveys} with an average sightline spacing of $\dperp = 2 \,
h^{-1}$Mpc and adjusted the volume to make problem sizes of $\log_2 \Npix = 15$,
16, 17, 18, and 19. We ran the $\log_2 \Npix = 15$, 16, and 17 problems with
OMP\_NUM\_THREADS set to 1, 2, 4, 6, 8, 12, 16, 18, and 24. We set the number of
threads to test the standard powers of two, but also included multiples of 6 to
test the Edison NUMA boundaries, which have a significant effect.

In the first panel of Figure~\ref{fig:scaling}, we show the strong scaling
efficiency (the walltime compared to what is expected for linear scaling)
for these problems. We show the efficiency $e_{\rm strong} = t_n / (n t_1)$
where $n$ is the number of threads and the $t$ is the walltime for each run.
The result is independent of the problem size. The relative speedup drops from
1.0 to 0.87 as the number of threads increases from 1 to 6, and then remains the
same up to a full node. This suggests there is an increasing (but small) cost
for threads to access memory until we hit the first NUMA barrier at 6 threads
and is constant after that. In the middle panel, we show a weak scaling problem,
increasing the number of threads from 1 to 4 to 16 as the problem size doubles
(since the algorithm scales as $\Npix^2$). We show the efficiency
$e_{\rm weak} = t_n / t_1$.
The decrease in efficiency is similar to the strong scaling case, where the 16
thread case is 0.88 of the max efficiency. Finally, in the third panel we show
the walltime per $\Npix^2$ element from problems all run with 24 threads,
doubling in size. The up-down pattern in this panel is not due to random system
behavior, but instead the number of PCG iterations. The bottom runs took 12
iterations while the top took 13 due to small differences in the noise
realizations. This test confirms the expected $\Npix^2$ scaling of the code and
also demonstrates how fast the code is. If we consider the number of threads
$n = 24$, the number of iterations $i = 12$, the clock speed
$s = 2.4 \, \mathrm{ns}^{-1}$, and the element-wise time $t = 3.2 \, \mathrm{ns}$,
the number of clock cycles taken per element per iteration is $n s t / i = 15$.
This is close to our estimate of 10 operations per lookup and multiply, even
though the element-wise time measurement is an overestimate, including other
operations like the $\vSmp \vx$ multiply.

\subsection{Error estimation}

There are two possibilities for estimating the errors of the map values. First,
we can compute the map covariance $\vM = \vSmp (\vSpp + \vN)^{-1} \vSpm$
directly. This option is straightforward, but prohibitively expensive
computationally. The inverse and product on the right of the map covariance is
now a matrix instead of a vector, meaning we must run a PCG solve for each row
of the solution matrix. One could also abandon an iterative method and perform a
direct inverse. Either way, the computational complexity of the covariance
calculation is a factor of $\Npix$ greater than the map calculation. For any
interesting problem, this is very expensive indeed.

Instead, we propose using Monte Carlo error estimation. We run $n$
reconstructions on data with random noise realizations (consistent with the
noise estimates), and estimate the map variance over the $n$ results. We expect
the required number of reconstructions $n$ to be much smaller than $\Npix$,
making this method much cheaper. For synthetic data sets, such as in this work,
this method also allows us to test the effect of noise in the data and the
effect of the sightline sampling independently.

\subsection{Alternate smooth map construction}

For our protocluster application, we are primarily interested in large-scale
fluctuations. A simple way to pick out large-scale fluctuations is to smooth the
field on the scale we are interested in, as we did earlier. This acts as a basic
matched filter. However, instead of smoothing a high-resolution reconstruction,
we could start with a different estimator that picks out large-scale
fluctuations. We can think of our signal split into low and high-frequency
components $\vs = \vs_l + \vs_h$. The Wiener Filter estimate of the low-frequency
signal is $\hat{\vs}_l = (\vS_{ll} + \vS_{hl}) (\vS + \vN)^{-1} \vd$.
We can split the signal with a Gaussian filter $\vG$ such that $\vs_l = \vG \vs$
and $\vs_h = \vs - \vG \vs$. It follows that $\hat{\vs}_l = \vS \vG (\vS +
\vN)^{-1} \vd$. Compare this to the expression for a smoothed map, $\vG \hat{\vs}
= \vG \vS (\vS + \vN)^{-1} \vd$. These expressions only
differ by the position of the Gaussian filter, but it is an important
distinction. In the case of the smoothed map, the filter acts on the map values,
whereas in the case of the smooth signal reconstruction, the filter acts on the
weighted pixel values. However, for any practical case where the filter scale is
larger than the pixel and map spacing, these expressions will be very close to
one another, and the distinction is no longer important.

\section{Optimal Filter for Protocluster Signal}
\label{app:opt_filt}

In this section, we explain a more advanced procedure for identifying
protoclusters in the flux maps. We exploit the fact that we know the shape of
the protocluster signal. We assume the flux map $d(\vx)$ is a combination of the
protocluster signal and  the background fluctuations of the Ly$\alpha$ forest.
That is, $d(\vx) = A \tau(\vx) + \delta_F(\vx)$, where $\tau(\vx)$ is the shape
of the protocluster profile and $A$ is the strength of this signal.

In this case, the derivation of the optimal filter is shown in Appendix A of
\citet{haehnelt_and_tegmark_1996}, which we briefly review. We estimate the
protocluster signal by convolving with a filter $\psi(\vx)$, so that $\hat A =
\int \psi(\vx) d(\vx) d^3x$. The filter is normalized such that the estimate is
unbiased, requiring $\int \psi(\vx) \tau(\vx) d^3x = 1$. In Fourier space, the
unbiased, minimum variance estimator is then $\tilde \psi(\vk) = C \tilde
\tau(\vk) / P(\vk)$, where tildes indicate the Fourier transform of a quantity,
$P(\vk)$ is the power spectrum of $\delta_F(\vx)$, and $C$ is the normalization
constant.

\begin{figure}
  \begin{center}
    \resizebox{3in}{!}{\includegraphics{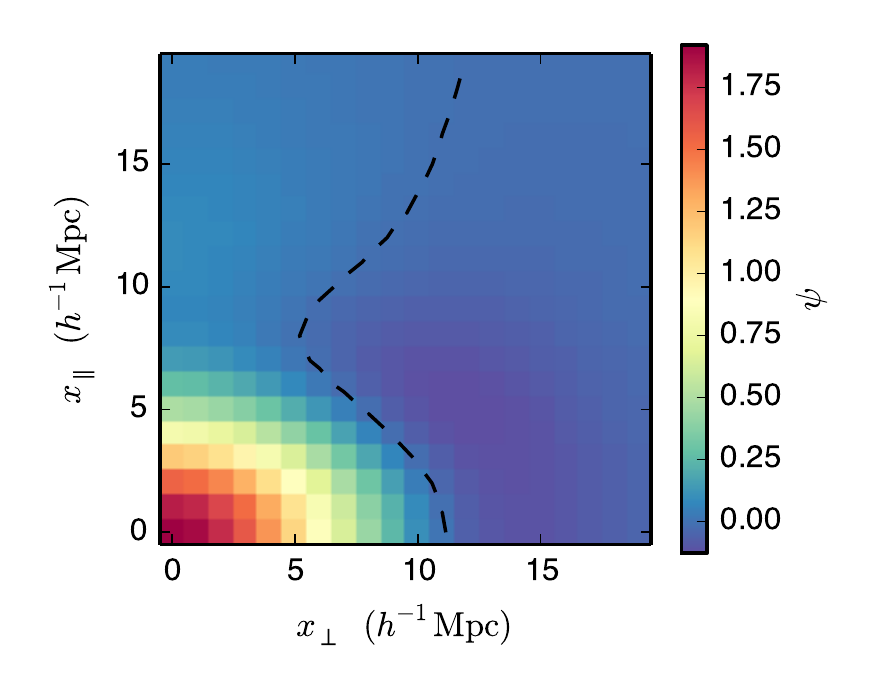}}
  \end{center}
  \caption{The optimal filter $\psi$ in the $x_\perp, x_\parallel$ plane. The
  black dashed line shows the $\psi = 0$ contour.}
  \label{fig:opt_filt}
\end{figure}

The optimal filter requires models for the protocluster profile and Ly$\alpha$
forest power spectrum. We model the protocluster profile as a Gaussian product
parallel and perpendicular to the line of sight. This is similar to the data and
model shown in Figure~\ref{fig:pc_profiles}, although in this case we break
spherical symmetry into the perpendicular and parallel components. We found that
the average protocluster has a Gaussian $\sigma$ scale of about $7 \, h^{-1}$Mpc
perpendicular to the line of sight, and is reduced to about $4 \,
h^{-1}$Mpc along the line of sight due to redshift-space distortions. We fit the
Ly$\alpha$ forest power spectrum with a Kaiser and isotropic Gaussian-damped
redshift-space power spectrum model, $P(k_\perp, k_\parallel) = a k^\alpha
(1 + \beta k_\parallel^2 / k^2)^2 \exp(-k^2 \sigma^2)$.
The normalization of the power is
set by a combination of the bias of the Ly$\alpha$ forest the normalization of
the primordial power spectrum. The $k^\alpha$ term accounts for a simple form of
the primordial power spectrum scaling, which is sufficient for the scales in the
simulation. The Kaiser term $(1 + \beta k_\parallel^2 / k^2)$ handles the
effects of redshift-space distortions on large scales. We include an isotropic,
Gaussian
damping term in order to capture suppression of small-scale fluctuations either
due to pressure support or the smoothing effect of the Wiener filter. We found
that the values $\alpha = -1.85$, $\beta = 1.07$, $\sigma = 2.06 \, h^{-1}$Mpc
provided a good fit. The resulting filter, in configuration space, is shown
Figure~\ref{fig:opt_filt}. It is encouraging to see a negative region in the
plot of $\psi(x_\perp, x_\parallel)$. This means that the filter will naturally
downweight modes which are dominated by background Ly$\alpha$ forest
fluctuations. This is an improved filter compared to the 3D Gaussian filter used
in the rest of the text, but we found that it did not make a significant
difference in the candidate identification result.

\end{document}